%% file: 6308.tex
\documentclass{aa}

\usepackage{graphicx}
\usepackage{amsfonts}
\usepackage{amsmath}
\usepackage{longtable}
\usepackage{txfonts}

\newcommand{\msun}{M_{\odot}}
\newcommand{\bd}{brown dwarf }
\newcommand{\bds}{brown dwarfs }
\newcommand{\bl}{Blanco~1 }

\newcommand{\oversim}[2]{\protect{\mbox{\lower0.5ex\vbox{%
   \baselineskip=0pt\lineskip=0.2ex
   \ialign{$\mathsurround=0pt #1\hfil##\hfil$\crcr#2\crcr\sim\crcr}}}}}
\newcommand{\simless} {\mbox{$\,\mathrel{\mathpalette\oversim<}\,$}} 

\begin{document}

  \title{The lower mass function of the young open cluster Blanco~1:
  from $30 M_{Jup}$ to $3 \msun$}

  \author{E. Moraux\inst{1}\thanks{Based on observations obtained at
    the Canada-France-Hawaii Telescope (CFHT) which is operated by the
    National Research Council of Canada, the Institut National des
    Sciences de l'Univers of the Centre National de la Recherche
    Scientifique of France, and the University of Hawaii. Based on
    observations collected at the European Southern Observatory, Chile
    (ESO programmes 68.C-0233 and 71.C-0446). Some of the data
    presented herein were obtained at the W.M. Keck Observatory, which
    is operated as a scientific partnership among the California
    Institute of Technology, the University of California and the
    National Aeronautics and Space Administration.}\and
    J. Bouvier\inst{1}\and J.R. Stauffer\inst{2}\and D. Barrado y
    Navascu\'es\inst{3} \and J-C. Cuillandre\inst{4,5}}

  \offprints{Estelle.Moraux@obs.ujf-grenoble.fr}


  \institute{Laboratoire d'Astrophysique, Observatoire de Grenoble
    (LAOG), B.P.  53, 38041 Grenoble Cedex 9, France
    \and Spitzer Science Center, Caltech M/S 220-6, 1200 East
    California Boulevard, Pasadena, CA 91125, USA
    \and Laboratorio de Astrofisica Espacial y Fisica Fundamental
    (LAEFF-INTA), Apdo. 50727, 28080 Madrid, Spain
    \and Canada-France-Hawaii Telescope Corporation, 65-1238 Mamalahoa
    Highway, Kamuela, HI 96743, USA
    \and Observatoire de Paris, 61 Av. de l'Observatoire, 75014 Paris,
    France}

  \date{Received 14/09/2006; Accepted }

  \authorrunning{E. Moraux et al.}

  \titlerunning{Low mass stars and brown dwarfs in the young open
  cluster Blanco~1}


  \abstract 
  {} 
  {We performed a deep wide field optical survey of the young ($\sim
  100-150$ Myr) open cluster Blanco~1 to study its low mass population well
  down into the brown dwarf regime and estimate its mass function over
  the whole cluster mass range.}
  {The survey covers 2.3 square degrees in the $I$ and $z$-bands down
  to $I\simeq z\simeq 24$ with the CFH12K camera. Considering two
  different cluster ages (100 and 150 Myr), we selected cluster member
  candidates on the basis of their location in the ($I,I-z$) CMD
  relative to the isochrones, and estimated the contamination by
  foreground late-type field dwarfs using statistical arguments,
  infrared photometry and low-resolution optical spectroscopy.}
  {We find that our survey should contain about 57\% of the cluster
  members in the $0.03-0.6\msun$ mass range, including 30-40 brown
  dwarfs. The candidate's radial distribution presents evidence that
  mass segregation has already occured in the cluster. We took it into
  account to estimate the cluster mass function across the
  stellar/substellar boundary. We find that, between $0.03\msun$ and
  $0.6\msun$, the cluster mass distribution does not depend much on
  its exact age, and is well represented by a single power-law, with
  an index $\alpha=0.69\pm 0.15$. Over the whole mass domain, from
  0.03$\msun$ to 3$\msun$, the mass function is better fitted by a
  log-normal function with $m_{0}=0.36\pm 0.07\msun$ and
  $\sigma=0.58\pm0.06$.}
  {Comparison between the Blanco~1 mass function, other young open
  clusters' MF, and the galactic disc MF suggests that the IMF, from
  the substellar domain to the higher mass part, does not depend much
  on initial conditions. We discuss the implications of this result on
  theories developed to date to explain the origin of the mass
  distribution.}


  \maketitle


\section{Introduction}

Since the first brown dwarf discovery and confirmation in 1995
(Nakajima et al. 1995; Rebolo et al. 1995), new perspectives have
opened regarding the formation of condensed objects in molecular
clouds. Today more than a thousand brown dwarfs (BDs) are known but
their mode of formation is still controversial and the theoretical
framework describing the stellar and substellar formation process(es)
is far from being satisfactory. How do brown dwarfs form~? Is there a
lower mass limit for an object to be formed~? A way to tackle these
questions is to determine the mass spectrum resulting from the stellar
formation process, i.e. the initial mass function (IMF), down to the
substellar regime.

Young nearby open clusters are ideal environments for such a
purpose. Their members constitute a uniform population in terms of
distance and age, the extinction is usually low and uniform, and their
youth ensures that brown dwarfs are still bright enough to be
easily detected and followed-up. Moreover, the well-known stellar
population of the nearest open clusters complements the discoveries of
brown dwarf members to yield a complete mass function from substellar
objects up to massive stars.

In this contribution we present a study of Blanco~1, sometimes
referred to by the name of its brightest member $\zeta$ Sculptoris.
It is a young southern open cluster ($\alpha_{J2000}=0^{\rm h}04^{\rm
m}24^{\rm s}$, $\delta_{J2000}=-29\degr56.4\arcmin$; Lyng\aa 1987)
located at an Hipparcos distance of $260^{+50}_{-40}$ pc (Robichon et
al. 1999) at a high galactic latitude (some 240 pc below the galactic
plane, $b=-79.3\degr$), and has a low extinction ($E(B-V)=0.010$). It
has a low stellar density with $\sim 200$ known stellar members spread
over a $1.5\degr$ diameter surface (Hawkins \& Favata 1998) and
its age is not very well constrained, around 100-150 Myr. Panagi \&
O'Dell (1997) found it is similar to or slighty older than that of the
well-studied Pleiades cluster of 120 Myr, while other authors used an
age of 100 Myr (e.g. Pillitteri et al. 2003).

The cluster's youth combined with its large distance from the galactic
disc point to an unusual formation history. Ford, Jeffries \& Smalley
(2005) found also that the abundance pattern in Blanco 1 is quite
exceptional with an average [Fe/H] close to solar ($=+0.04\pm0.04$)
but a subsolar [Ni/Fe], [Si/Fe], [Mg/Fe] and [Ca/Fe]. This lead the
authors to suggest that the material from which the cluster formed was
different from the local galactic disc ISM. It may have travelled some
distance without having the chance to homogenize with the ISM and may
also have been polluted by one or two unusual supernova events. All
these peculiarities make \bl a very interesting target to test the
dependence of the IMF on environmental conditions when comparing its
present day mass function (MF) to those of similar age open clusters.

\bl was discovered by Blanco (1949) who noticed a small concentration
of A0 stars having the same galactic latitude in this region. Since
then several photometric studies have been performed, e.g. by
Westerlund (1963), Epstein (1968), Eggen (1972), Perry et al. (1978),
Lyng{\aa} \& Wramdemark (1984), Abraham de Epstein \& Epstein (1985)
and Westerlund et al. (1988). Abraham de Epstein \& Epstein (1985)
published a relatively complete list of 260 F, G, and K stellar
candidates. Kinematic studies based on radial velocities (Jeffries and
James 1999) and proper motions (Pillitteri et al 2003) indicate that
the contamination level of this photometric sample is about 35 to
40\%. X-ray observations performed by Micela et al. (1999) yielded the
detection of a few lower mass sources -- M dwarfs -- confirmed by
proper motion. However, the \bl population at very low masses and in
the substellar domain is still unexplored.

We performed a deep wide-field optical survey and follow-up
observations to look for \bl brown dwarfs (BD) and very low mass stars
(VLM) in order to estimate its mass function (MF) down to the
substellar domain. The observations and data reduction are described
in Section 2 and our results are given in Section 3. In Section 4 we
compare our MF estimate to the Pleiades, other young clusters and the
galactic disc MF to investigate the IMF universality. We discuss the
implications for recently developed IMF theories.


\section{Observations and data reduction}

\subsection{Optical survey}

\begin{figure}[htbp]
  \includegraphics[width=0.9\hsize]{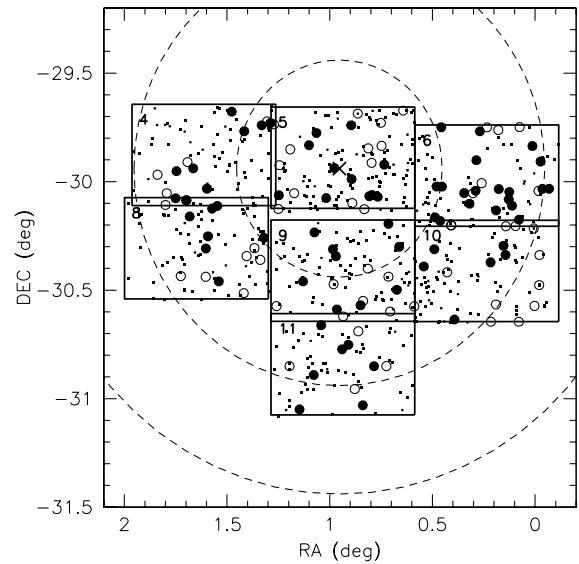}
  \caption{Area of the sky covered by the CFHT 1999 and 2000 Blanco~1
    surveys. Each rectangle corresponds to one CFH12K field (the
    coordinates are given in Table~\ref{fields}). The small cross in
    the middle represents the nominal cluster center and the dashed
    circles have a radius of 0.5, 1 and 1.5 degrees
    respectively. Small dots show the location of low mass star
    candidates selected from the short exposures. Very low mass star
    and brown dwarf candidates selected from the long exposures are
    shown as large dots if confirmed by infrared photometry and/or
    optical spectroscopy (open circles if ruled out).}
  \label{bl_survey}
\end{figure}

\bl has been observed using CFHT's 12k$\times$8k optical CCD mosaic
camera (Cuillandre et al. 2001) during two separate runs in September
1999 and December 2000. Over the two runs, a total of 7 non
overlapping fields, each of size $28\arcmin\times42\arcmin$ on the
sky, were obtained in the $I$ and $z$ filters (central wavelengths
$\sim850$ and 950 nm respectively). The covered fields are graphically
shown in Fig.~\ref{bl_survey} and their coordinates are listed in
Table~\ref{fields}. For each pointing and each filter, the observing
sequence included one 10s and two 600s exposures. A total area of 2.3
sq.deg. was thus surveyed down to a detection limit of $I\sim z\sim24$
on and around the cluster's nominal center. The seeing measured on
images obtained in both filters spanned a range from 0.5 to 0.8 arcsec
FWHM.

\begin{table}[htbp]
  \centering
  \caption{J2000 coordinates of the 7 CFH12K fields covered in
  Blanco~1.}
  \input{6308tab1.tex}
  \label{fields}
\end{table}

These observations have been performed with the same instrument and
partly during the same nights as the Pleiades cluster survey described
by Moraux et al. (2003). The reduction of the CFH12K mosaic images has
been done likewise, chip by chip. Briefly, they were first overscan
and bias subtracted and a dark map correction was applied. Supersky
flats were computed combining all images obtained in each band during
the whole course of the observing run, and they were normalized
to a reference CCD in order to obtain a uniform gain over the whole
mosaic. The same photometric zero-point can thus be used for all the
CCDs. Images were then flat-field corrected and fringes were removed
(see Magnier \& Cuillandre 2004 for details of the reduction
procedure). Then the pre-reduced pairs of 600s exposures of each field
were finally combined to yield the final long exposure images in the
$I$ and $z$ filters.

The photometric analysis was done using the SExtractor package (Bertin
\& Arnouts 1996). To detect the maximum number of sources we first
added all pairs of $I$ and $z$ reduced images together and we used the
automatic object-finding algorithm on these frames. We then computed
the photometry of each detected object by PSF-fitting on the
individual (i.e. $I$ or $z$) images. Spurious detections (cosmic rays,
bad columns) and extended sources were rejected on the basis of their
FWHM. For both runs, 1999 and 2000, the zero point of the instrument
in the $I$-band was derived using Landolt photometric standard fields
(SA98 and SA113) observed in the same conditions and reduced in the
same way as \bl fields. Several exposures of SA98 were obtained
with an incremental offset of several arcminutes in RA so that common
sets of standard stars were observed on every CCD of the mosaic. We
thus checked that the photometric zero-point in the $I$-band was the
same for each CCD of the mosaic. For the $z$ filter we used
unreddened A0 standard stars covered by the Landolt fields assuming
their $I$ and $z$ magnitudes were the same. For $I<20$, the resulting
photometric rms ranges from 0.03 to 0.05 mag for both $I$ and $z$
filters, which corresponds to a colour accuracy between 0.04 and
0.07 mag in $I-z$. For fainter objects the accuracy is slighty worse
but remains better than 0.07 mag in $I$ and $z$, or 0.10 mag in $I-z$,
down to $I\sim22.5$.

The astrometric solution of each image was calculated with the Elixir
software (Magnier \& Cuillandre 2002). We first computed it for the
short exposures using the USNO2 catalog as reference and we used this
solution as a first approximation for the long exposures. Since most
of the USNO2 stars were saturated in the 600s images we could not use
this catalog and we used the stars detected in both short and long
exposures as astrometric references instead. The same technique was
used by Moraux et al. (2003) and the resulting astrometric accuracy is
about 0.3 arcsec which is good considering the CFH12K plate scale
of 0.206 arcsec/pixel. 

\subsection{Infrared follow-up photometry}

We observed in the near-infrared 90 of the 124 brown dwarf and very
low mass star candidates selected from the long exposures of the
optical survey (see section~\ref{bdselect}) to confirm their status.

We obtained $K_{s}$-band images for 77 of them using the SOFI camera
(1024$^2$ pixels, 0.288 arcsec/pixel) at ESO/NTT in August 2003. The
total exposure time per object was in the range from 10 to 20 min,
from the brightest ($K_{s}\sim16$) to the faintest source
($K_{s}\sim20$). The seeing was typically around 0.8 arcsec during the
entire run and the conditions were most of the time photometric.

Thirteen other candidates have been observed with the near-infrared
camera CFHT-IR (1024$^2$ pixels, 0.211 arcec/pixel) in November
2004. The observing conditions were very good, photometric with a
seeing of about 0.6 arcsec.

In both cases, individual infrared images obtained for each object at
7 or 9 dithered positions were first median combined in order to
obtain a sky image which was subtracted from the original images. Sky
subtracted images were then flat field corrected using dome flats and
averaged to yield the final $K_{s}$ image of the object. Aperture
photometry was performed on the object using the IRAF/APPHOT
package. Several photometric standards from Persson et al. (1998) were
observed each night throughout the run and reduced in the same way as
the \bl candidates. They were used to derive the $K_{s}$-band
extinction coefficient
as well as the zero-point of the instrument (ZP $=22.46\pm 0.02$ mag
for SOFI and ZP$=22.73\pm 0.01$ for CFHT-IR) and the final photometry
is given in the $K$-band Las Campanas system. The resulting internal
photometric precision on the $K$-band measurements of \bl candidates
is typically 0.02-0.03 mag for $K\le 18$ and 0.05-0.1 mag for fainter
objects. While most of \bl targets were observed during photometric
conditions, a few (BL-6, -70, -85, -102, -111) were affected by thin
cirrus at the end of the 4th and 5th nights of the SOFI run. In order
to calibrate the images taken in such conditions, we re-obtained
shorter exposure time images of these fields during the following
photometric nights. For those few candidates, the photometric accuracy
is not as good, of order of 0.10 mag.

\subsection{Optical spectroscopy}

In addition to the photometric data we obtained low resolution optical
spectroscopy for the brightest \bl \bd candidates. The observations were
performed using FORS2 ($R\sim1500$) at the VLT in November 2001 and
LRIS ($R\sim1000$) at Keck in December 2001 and October 2002. For the
three runs the data reduction was done in the same way using the IRAF
package. Briefly, the CCD images were overscan, bias and flat-field
corrected, the sky background was subtracted, and the spectra were
extracted using the command APALL. The spectra were then wavelength
calibrated using HgNeAr spectra and corrected from the instrument
response computed by dividing observed standard star spectra by their
reference spectra.


\section{Results}

In the following we present the results we obtained from the analysis
of the short (Sect. 3.1) and long (Sect. 3.2) optical exposure
images. The corresponding detection limits are $I\sim13-21$ for
$t_{int}=10$s and $I\sim17.5-24$ for $t_{int}=600$s which allow our
study to be complete from $I=14$ to $I=22.2$. For both integration
times, we constructed the ($I$, $I-z$) colour-magnitude diagram (CMD)
from which we selected low mass star and brown dwarf candidates.  In
order to take into account the uncertainty on the cluster age we made
two selections, one for an age of 100 Myr and the other for 150
Myr. For each sample, we estimate the contamination by field dwarfs
using either statistical arguments or additional near-infrared
photometry and optical spectroscopy when available. We then
investigate the cluster member spatial distribution and determine the
cluster core radius for different mass ranges (Sect. 3.3). We finally
use these results to derive the Blanco1 mass function from 30 Jupiter
masses to $3\msun$ (Sect. 3.4).

\subsection{Low-mass candidate selection ($0.09\msun\le m\le0.6\msun$)}

\begin{figure}[htbp]
  \includegraphics[width=0.9\hsize]{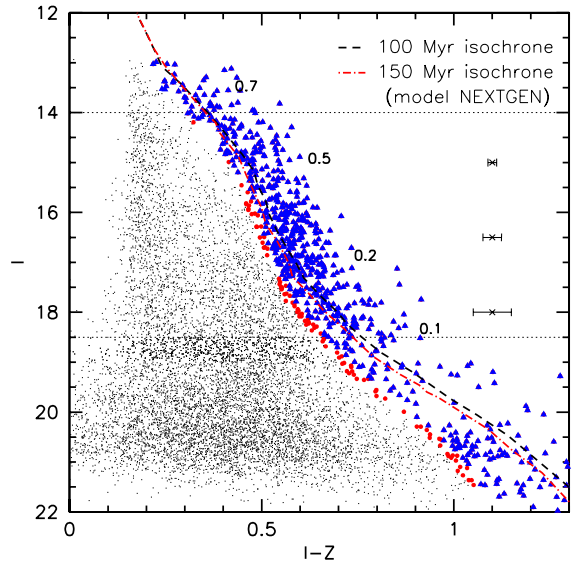}
  \caption{($I,I-z$) colour-magnitude diagram for the short exposures
    of the optical survey. The dashed (dot-dashed) line is the 100 Myr
    (resp. 150 Myr) isochrone from the NEXTGEN models of Baraffe et
    al. (1998) shifted to the \bl distance. The triangles and large
    dots correspond to our photometric selection of cluster member
    candidates for both ages (see text). Typical error bars
    are shown on the right side.}
  \label{cmdshort}
\end{figure}

The ($I$, $I-z$) CMD corresponding to the short exposures of the
optical survey is presented Fig.~\ref{cmdshort}. The 100 Myr and 150
Myr isochrones given by the NEXTGEN models from Baraffe et
al. (1998)\footnote{The NEXTGEN and DUSTY isochrones have been
calculated especially for the CFH12K filter set by I.Baraffe using the
transmission curves and the CCD quantum efficiency. Their reliability
is ensured by the fact that they work well at reproducing the Pleiades
single star sequence for an age of 120 Myr and a distance of 125 pc in
agreement with the cluster properties} (see Moraux et
al. 2003). shifted to the cluster distance ($(m-M)_{o}= 7.1\pm 0.4$;
Robichon et al. 1999) are shown as a dashed and dot-dashed line. In
order to avoid incompleteness due to detection bias (either saturation
on the bright side or too low signal to noise ratio on the faint
objects) we restricted our study to the magnitude range from $I=14$ to
18.5. We then made a rather conservative photometric selection
for each age (100 Myr and 150 Myr) by shifting the corresponding
NEXTGEN isochrone to the maximum cluster distance given by Hipparcos
($(m-M)_{o}=7.5$), and on the blue side by the colour photometric
error given by:
\begin{equation}
 \sigma_{I-z}=0.014+2.10^{-8}\exp(0.8\times I)
 \label{sigmaIz}
\end{equation}
All the objects redder than this line are \bl candidate members. The
150 Myr isochrone is slightly bluer than the 100 Myr one, which means
that the second selection will pick up a few more sources. The objects
selected by both the 100 Myr and 150 Myr isochrones (by only the 150
Myr isochrone) are shown as triangles (resp. large dots) on
Fig.~\ref{cmdshort}. An electronic list of all the candidates is
available from the CDS website. According to the NEXTGEN model, the
magnitude range $I=14-18.5$ corresponds to a mass $m$ between
0.09$\msun$ and 0.6$\msun$ at 100 Myr ($0.10-0.6\msun$ at 150 Myr) for
$(m-M)_{o}=7.1$.

One of the major shortcomings of this photometric selection is the
contamination by sources unrelated to the cluster which may lie in the
same region of the CMD. While a fraction of the selected objects must
be true cluster members, others are merely field objects (galaxies,
red giants or older field stars) on the line of sight. We do not
expect to find many galaxies as contaminants because we rejected all
the extended objects through the photometric analysis. The number of
red giants in our sample is also expected to be small because they
would have to be at greater than 6 kpc above the galactic plane to
overlap in the CMD with Blanco1 members, and the density of giants at
that height is small. Instead most of the contaminants are expected
to be foreground field M-dwarfs.

\begin{figure}[!ht]
  \includegraphics[width=0.9\hsize]{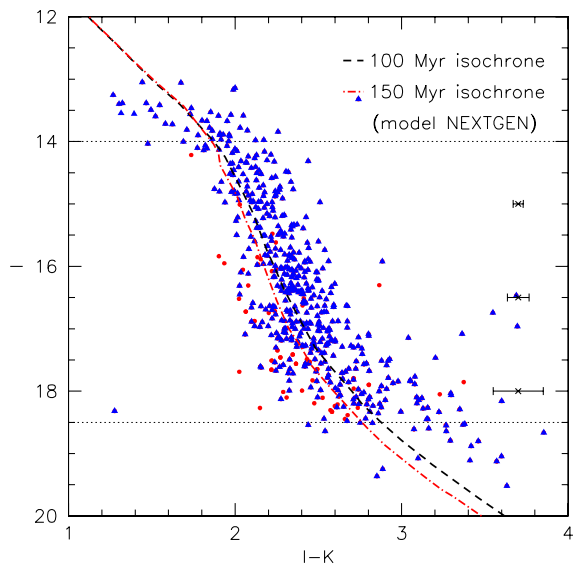}
  \caption{($I$, $I-K$) colour-magnitude diagram for all the
  optically selected candidates which have a 2MASS counterpart.  The
  dashed and dot-dashed line are the 100 Myr and 150 Myr isochrones
  from the NEXTGEN models of Baraffe et al. (1998) shifted to the \bl
  distance. The symbols are the same as in Figure~\ref{cmdshort}.}
  \label{cmdshort_ir}
\end{figure}

\begin{figure}[!ht]
  \includegraphics[width=0.9\hsize]{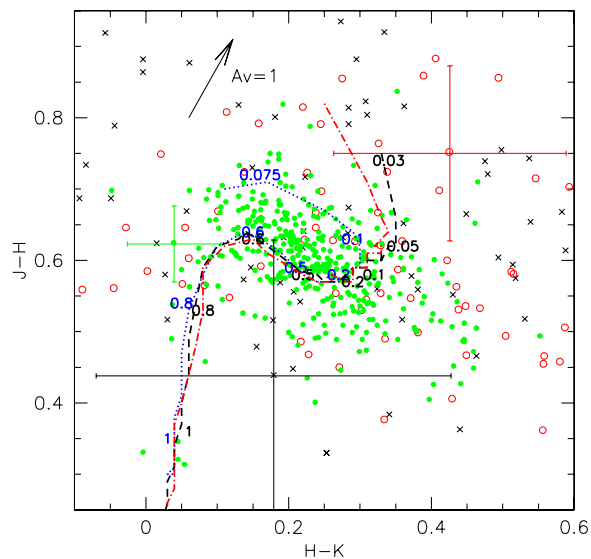}
  \caption{($J-H$, $H-K$) colour-colour diagram of all the low
  mass star candidates with a 2MASS counterpart. The different symbols
  correspond to different flags in the 2MASS catalog, i.e. different
  photometric accuracy (dots = flag A, open circles = flag B, crosses
  = flag C). A typical error bar is represented for each symbol. The
  100 Myr, 150 Myr and 5 Gyr (corresponding to the field dwarf locus)
  isochrones are shown as a dashed, dot-dashed and a dotted line
  respectively and are labelled with mass. The reddening vector is
  also shown. All candidates are located on or close to the isochrones
  -- within the error bars -- indicating that there is little
  extinction and that they all have near-infrared colours consistent
  with M-dwarf spectral type.}
  \label{jhhk}
\end{figure}

We cross-correlated the whole list of 534 candidates with the 2MASS
catalogue for $K\simless 15.5$. About 490 of them have a 2MASS
counterpart, the others are too faint. The corresponding ($I$, $I-K$)
CMD and ($J-H$, $H-K$) colour-colour diagram are shown in
Fig.~\ref{cmdshort_ir} and Fig.~\ref{jhhk} respectively. Most of the
candidates lie on or close to the 100 Myr or 150 Myr NEXTGEN isochrone
in both diagrams. This indicates that there is little extinction
towards Blanco~1 and that all the candidates have infrared colours
compatible with being M dwarfs. This supports our contention that
there are few giants or galaxies in our candidate sample. It shows
also that infrared colours fail to distinguish between field dwarfs
and cluster members in this mass range as expected from the
models.

To estimate the number of contaminants in our sample, we use instead
statistical arguments. For each ($I,I-z$) position in the CMD, we
first estimate the absolute $M_I$ magnitude of a field dwarf having
the same $I-z$ colour using the 5 Gyr NEXTGEN isochrone, and we derive
its distance from the difference $I-M_I$. An apparent magnitude range
$I,I+\Delta I$ yields a distance range and we compute the
corresponding volume containing the contaminants from the surface of
the sky covered by our survey. Using the M-dwarf $M_I$-band luminosity
function from Zheng et al. (2004) as well as their scale
height\footnote{Blanco~1 being at 240 pc below the galactic plane it
is indeed necessary to take the scale height into account.} we
calculate the number of expected contaminants per $M_I$ bin in the
magnitude range $I,I+\Delta I$. Taking into account a gaussian
photometric error of width $\sigma_{I-z}$ given by
equation~\ref{sigmaIz}, we then derive the number of objects per
$I-z$. Finally, to obtain the total number of field M dwarfs per
magnitude bin in our sample, we integrate over $I-z$, for $I-z$ redder
than our selection limit. The results are summarized
Table~\ref{contam}. Out of the 479 (534) candidates detected in our
survey for an assumed age of 100 Myr (150 Myr), we estimate about 211
(292) contaminants, which corresponds to a contamination level of
$\sim45\%$ ($\sim 55\%$).

\begin{table*}[htbp]
  \centering
  \caption{Estimated number of contaminants and probable cluster
  members for $14\le I\le 18.5$ for both ages 100 Myr and 150 Myr.}  
  \input{6308tab2.tex}

   \label{contam}
\end{table*}

\subsection{Brown dwarf and very low mass star selection ($0.03\msun\le
  m\le0.09\msun$) }

\label{bdselect}

\begin{figure*}[htbp]
  \includegraphics[width=0.9\hsize]{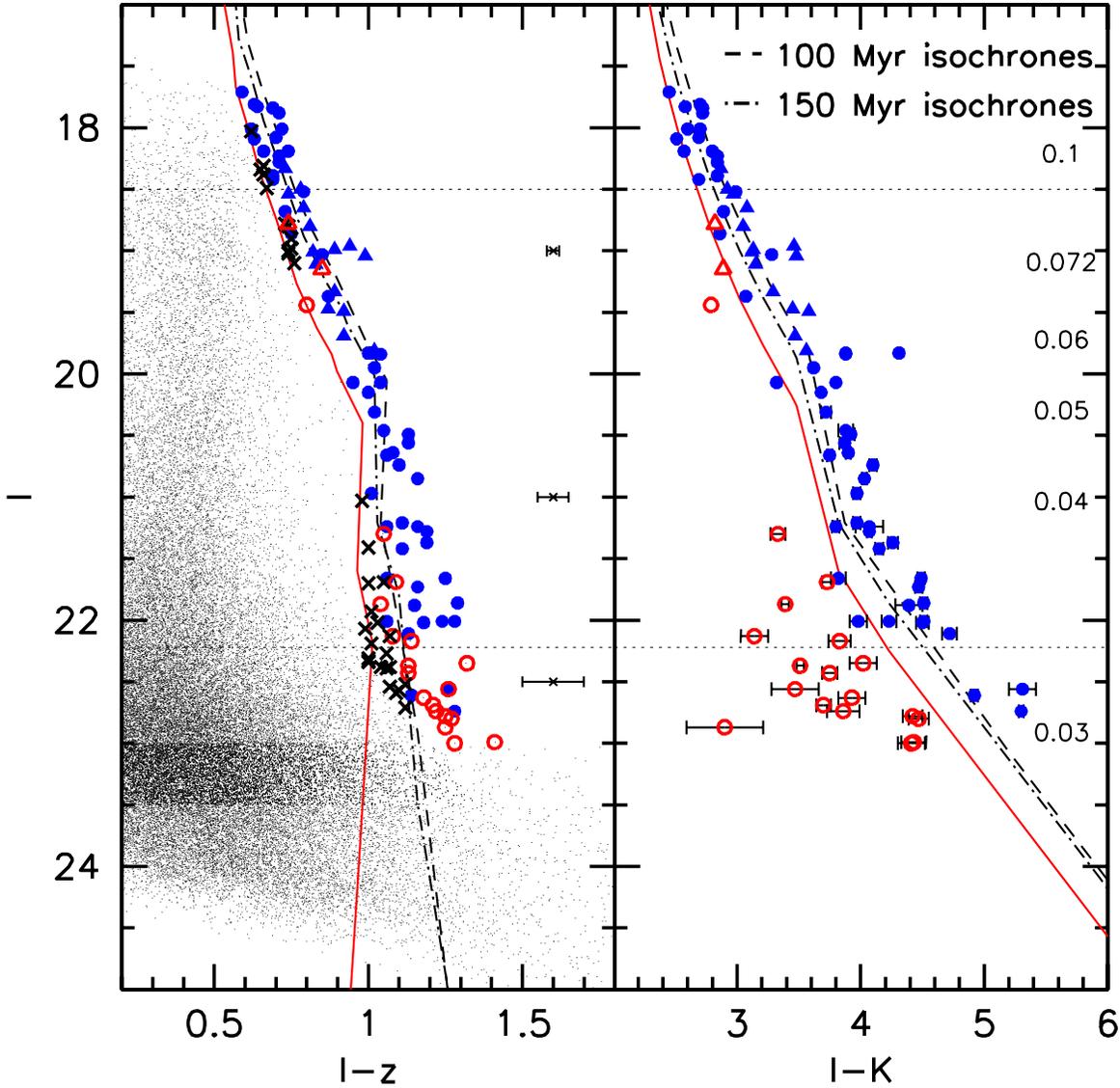}
  \caption{($I$,$I-z$) and ($I$,$I-K$) CMDs for all point-like
    objects identified in the long exposures of our survey. The 100
    Myr and 150 Myr isochrones constructed from the NEXTGEN and DUSTY
    models (see text) shifted to the distance of Blanco~1 are shown as
    a dashed line and dot-dashed line respectively. The mass scale (in
    $\msun$) corresponding to a cluster age of 100 Myr is given on the
    left hand-side of the ($I$,$I-K$) CMD. In each CMD, the solid
    line represents our bluest selection, i.e. for an age of 150 Myr,
    that maximizes the number of candidates. All the objects redder
    than this line on the ($I$, $I-z$) CMD have been identified as
    very low mass star and brown dwarf candidates. Sources with
    $I-K$ bluer than the solid line in the ($I$,$I-K$) CMD
    have then been rejected and are shown as open
    circles. Furthermore, candidates with optical spectroscopy
    consistent with cluster membership are shown as filled triangles
    (open triangles if not consistent).}
  \label{cmdlong}
\end{figure*}

   \begin{table*}[htbp]
     \centering
     \caption{Very low mass star and brown dwarf candidates selected
       from the long exposure images on the basis of their location in
       the ($I$,$I-z$) CMD (see Fig.~\ref{cmdlong}). Objects selected
       by the 150 Myr isochrone but not the 100 Myr isochrone are
       written in italic. An electronic version of this table can
       be found on CDS.}  \null\vskip -0.5cm \footnotesize
       \input{6308tab3.tex}

     \label{candcoo}
   \end{table*}

The optical ($I$,$I-z$) CMD corresponding to the long exposures of the
survey is shown on the left panel of Figure~\ref{cmdlong}, as well as
the 100 Myr and 150 Myr isochrones. These isochrones follow the
NEXTGEN models (Baraffe et al. 1998) down to $T_{eff}\simeq2500$~K,
i.e. down to $I\simeq 20.5$ for Blanco~1. Then dust starts to appear
in the \bd atmosphere and the NEXTGEN models are not valid any
more. When $T_{eff}\le 2000$~K ($I\simeq 21.5$), all the dust is
formed and the isochrones follow the DUSTY models (Chabrier et
al. 2000). In between there is a transition and we simply draw a
line from the NEXTGEN to the DUSTY models to ensure the continuity of
the isochrones.

To select the candidates we proceed in the same way as for the short
exposures. For each age, we shifted the isochrones vertically to the
largest cluster distance ($(m-M)_0=7.5$) and horizontally on the blue
side by the colour photometric error. All the objects redder than this
line are then considered as \bl VLM and BD candidates. This gives us
two samples depending on which age we use. The line corresponding to
the bluest selection (based on the 150 Myr isochrone) maximizes the
number of candidates and is shown on Figure~\ref{cmdlong}. We
identified 110 objects with $17.7\le I\le23$ for 100 Myr, and 14 more
for 150 Myr, i.e. 124 in total. They are listed in
Table~\ref{candcoo}. Their mass spans a range between $\sim0.03\msun$
and $0.1\msun$ and depends on the age of the cluster. Taking into
account the distance of the cluster, the stellar/substellar boundary
is around $I\simeq19.1$ at 100 Myr and 19.6 at 150 Myr according to
the NEXTGEN models, which means that about 75-80 of the selected objects
are \bd candidates.

\begin{figure}[htbp]
  \includegraphics[width=0.9\hsize]{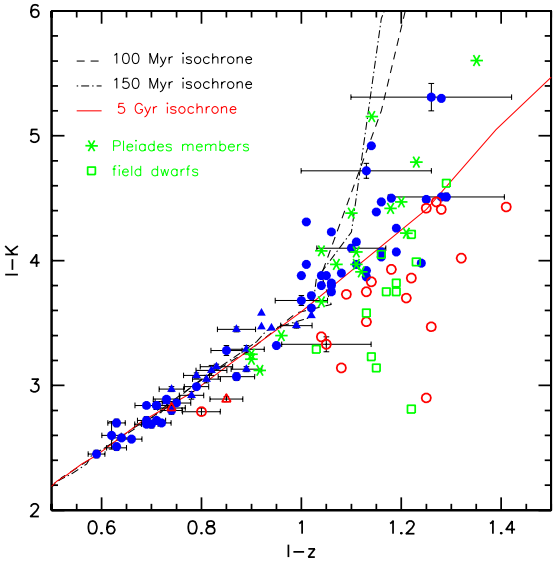}
  \caption{$(I-z,I-K)$ colour-colour diagram of the very low mass star
  and brown dwarf candidates. The symbols are the same as in
  Fig.~\ref{cmdlong}. The 100 Myr, 150 Myr and 5 Gyr isochrones are
  shown as a dashed line, dot-dashed line and solid line
  respectively. The asterisks (resp. open squares) represent Pleiades
  candidates confirmed (resp. rejected) by proper motion (Moraux et
  al. 2003, Casewell et al. 2007). Only a few error bars
  are drawn for clarity.}
  \label{Iz_IK}
\end{figure}

As explained in the previous section, this optical photometric sample
is contaminated by late M-type field dwarfs and we therefore conducted
follow-up observations of the candidates to confirm their
membership. We obtained near-infrared photometry with SOFI at ESO/NTT
for 77 of them, and with CFHT-IR for 13 others. The measured $K$-band
photometry is given in Table~\ref{candcoo}. The candidate location in
the ($I,I-K$) CMD (Fig.~\ref{cmdlong}, right panel) compared to the
isochrones shifted to $(m-M)_0=7.5$ allowed us to weed out 18 objects,
too blue for being Blanco~1 members. They are indicated as open
circles in Fig.~\ref{cmdlong}. The $I-K$ colour is a good diagnostic
of membership in the DUSTY regime, for $I\gtrsim21$, but not as much
for the brightest objects because field dwarfs and cluster members
have similar infrared colours for $I-z\lesssim1.0$ (see
Fig.~\ref{Iz_IK}). A statistical estimate using the Zheng et
al. (2004) luminosity function (LF) and calculated in the same manner
as for the short exposures but with
\begin{equation}
 \sigma_{I-z}=4.47.10^{-6}\exp(0.465\times I)
\end{equation}
indicates that there should be 10-15 contaminants between $I=17.7$ and
18.5. Assuming that we are about 40\% complete in this magnitude range
(because of saturation), this yields a contamination level of
$\sim30\%$. Below $I=18.5$, we estimate the survey is complete and we
find a contamination level of about 30\% for $I=18.5-20$. Between
$I=20$ and 21, we cannot use the same method because we reach the end
of the field M-dwarf LF but we do not expect the contamination to be
very high as the cluster sequence is well detached from the bunch of
field objects in the ($I,I-z$) CMD. The fact that we do not identify
any contaminant (except maybe CFHT-BL-57) in the ($I,I-K$) CMD in this
magnitude range confirms that the contamination level is indeed very
low. For $I\gtrsim21$ and $I-z>1.0$ infrared colour is sufficient to
assess membership. The $I-K$ colour of a 5 Gyr field dwarf starts
indeed to differ from that of a $\sim100$ Myr cluster member in this
domain (see Fig.~\ref{Iz_IK}). Even though the models may be too blue
in $I-z$, empirical data from Pleiades studies (Moraux et al. 2003,
Casewell et al. 2007) indicate that the $I-K$ colour difference is
real. Pleiades members confirmed by proper motion are redder in $I-K$
than older field dwarfs with the same $I-z>1.1$.

\begin{figure*}[htbp]
  \includegraphics[width=0.9\hsize]{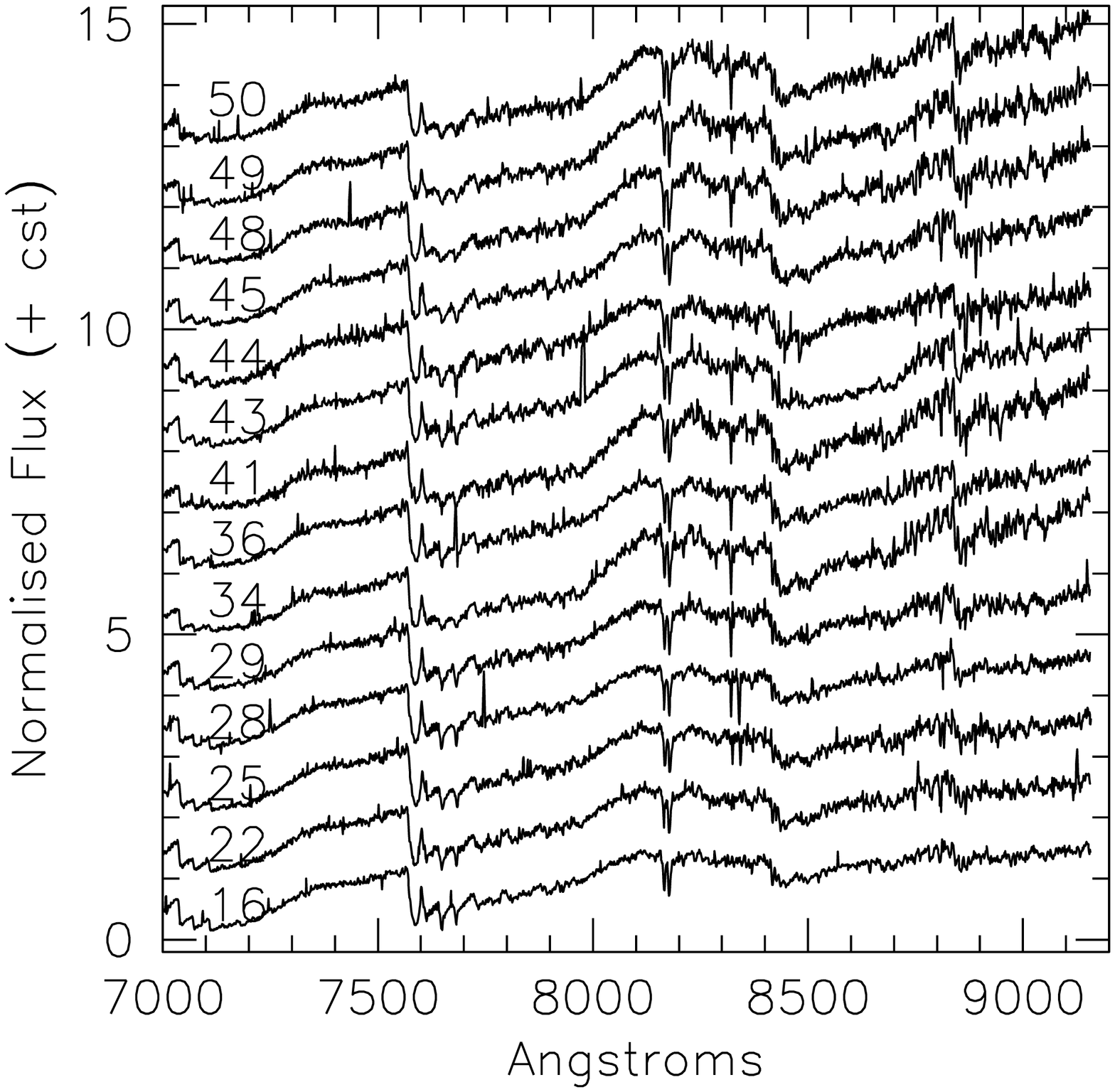}
  \caption{FORS2 spectra of CFHT-BL VLM candidates whose number is
  given above each spectrum on the left hand side.}
  \label{fors}
\end{figure*}

\begin{figure*}[htbp]
  \includegraphics[width=0.9\hsize]{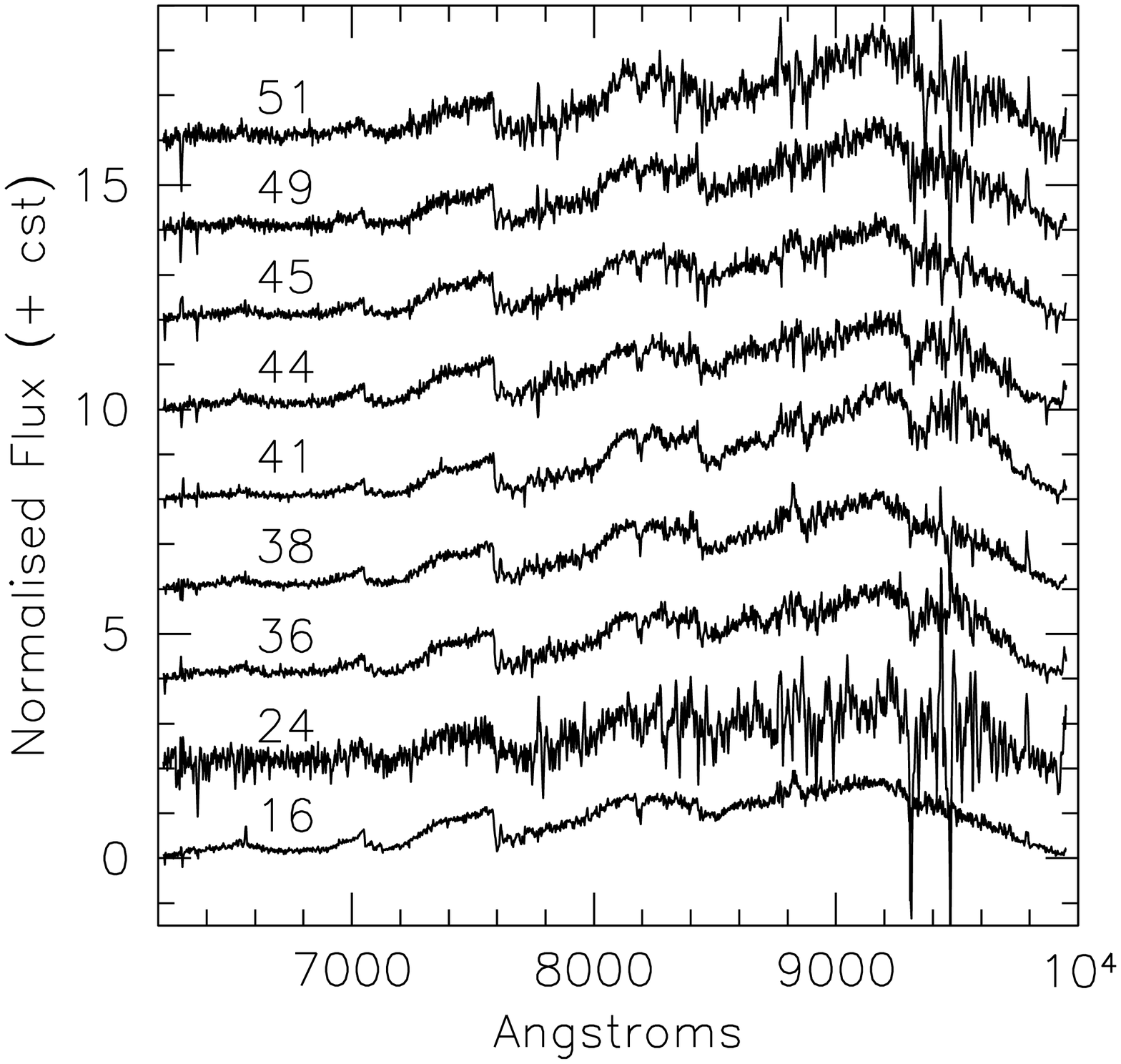}
  \caption{LRIS spectra of CFHT-BL VLM candidates.}
  \label{keck}
\end{figure*}

Additional spectroscopic data are thus especially needed for
candidates brighter than $I\sim20$. We obtained optical low resolution
spectroscopy for 17 of them with FORS2 on VLT and/or LRIS on Keck. The
spectra are shown Fig.~\ref{fors} and~\ref{keck}.


\begin{figure}[htbp]
  \includegraphics[width=0.9\hsize]{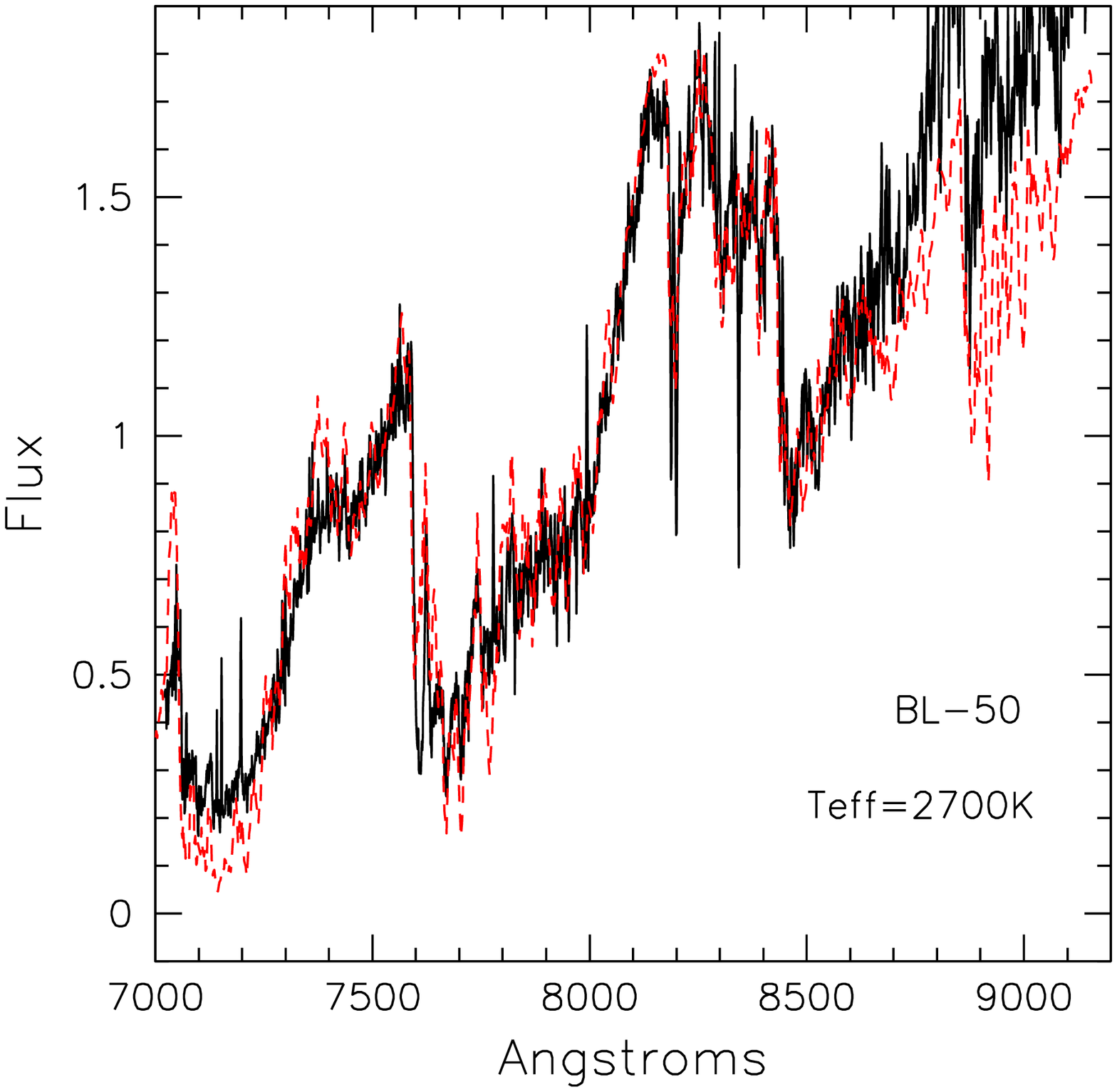}
  \caption{Spectrum of BL-50 (solid line) fitted by a synthetic
  spectrum (dashed line) from the DUSTY models of Allard et
  al. (2001). The best fit gives $T_{eff}=2700$K, which is in agreement
  with the M7-M7.5 spectral type we find using the PC3 index or the
  comparison to spectroscopic standards.}
  \label{fitTeff}
\end{figure}

From these data, we estimate the spectral type of each object using
different methods. We first use the PC3 index defined as the ratio of
the flux between 8230{\AA} and 8270{\AA} to the flux between 7540{\AA}
and 7580{\AA} in order to derive an estimate of the spectral type via
the relationship
\begin{equation*}
SpT = -6.685 + 11.715 \times (PC3) - 2.024 \times (PC3)^2
\end{equation*}
given by Martin et al. (1999). We also compare our spectra to those of
spectroscopic standard stars obtained in the same conditions
which allows us to estimate the spectral type with a precision of one
sub-class. By fitting the data with synthetic spectra from the DUSTY
models by Allard et al. (2001), we also estimate the effective
temperature of our candidates with an accuracy of 100K
corresponding to the grid step. An example of such a fit is given in
Fig.~\ref{fitTeff} for BL-50. All these methods are in agreement
with each other and the results are given in Table~\ref{bl_spectro}.

    \begin{table*}[htbp]
      \centering
      \caption{Measurements made on the optical spectra of \bl
        candidates. In the few cases where an object has been observed
        with both FORS2 and LRIS, the two calculated PC3 values are
        given in column 2 and the spectral type given in column 3 has
        been derived using the average value. [A colon indicates
        uncertain detection due to low S/N ratio or blending.]}
        \null\hskip -0.7cm\small\input{6308tab4.tex}
      \label{bl_spectro}
    \end{table*}

\begin{figure}[htbp]
  \includegraphics[width=0.9\hsize]{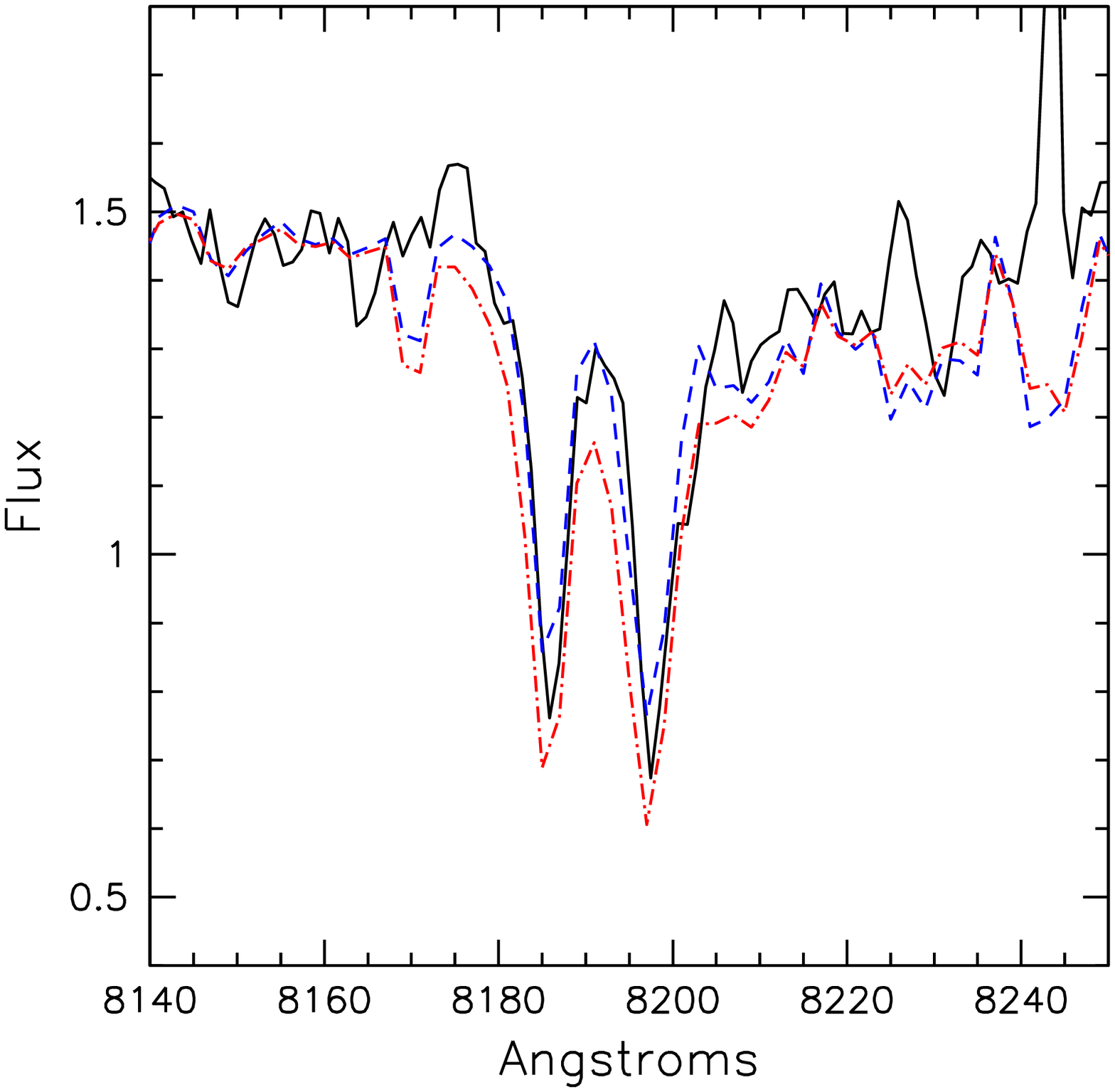}
  \includegraphics[width=0.9\hsize]{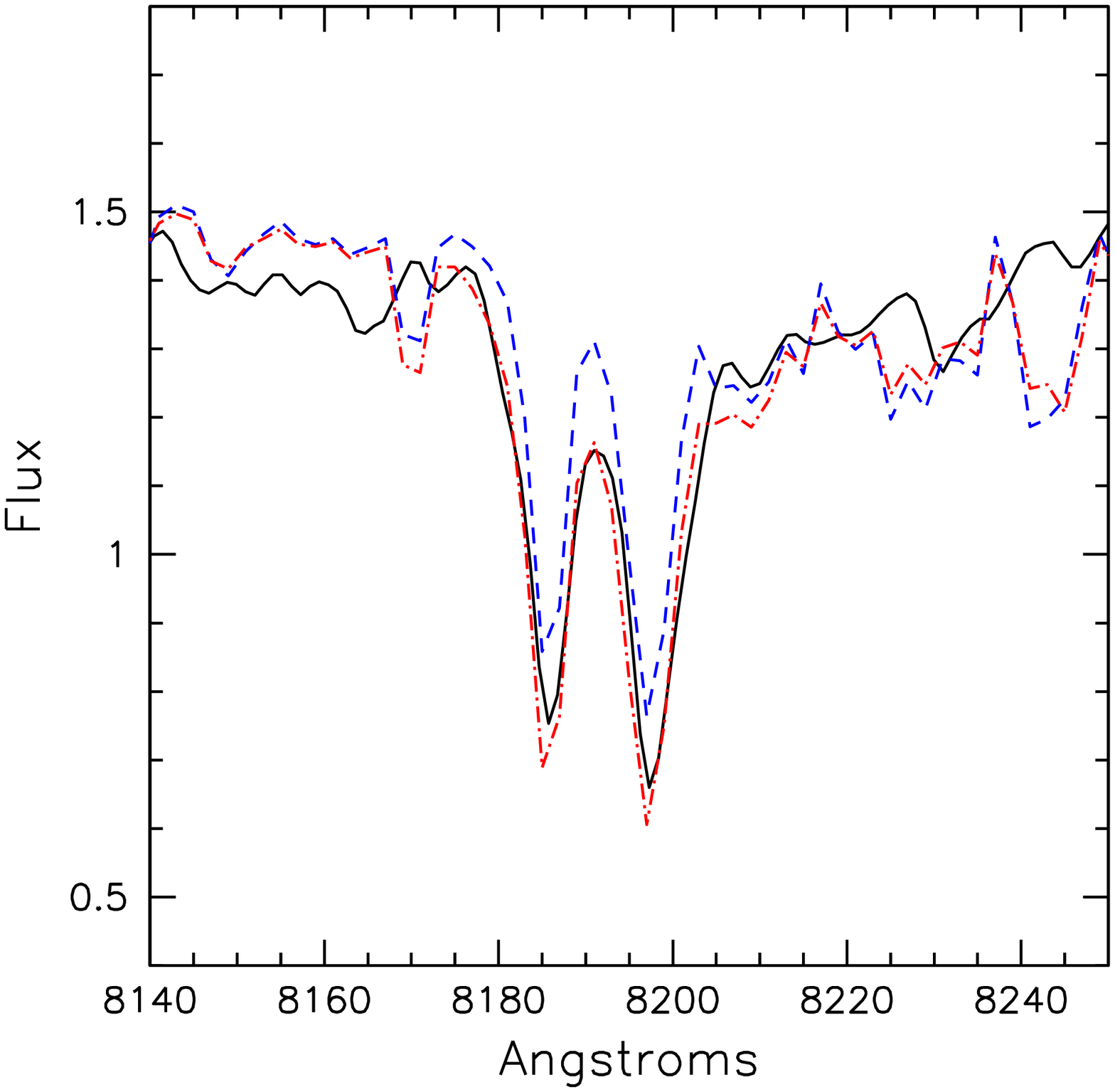}
  \caption{Upper panel: Comparison of the Na~I doublet profile of
  BL-29 (solid line) with two synthetic spectra of similar $T_{eff}$
  for $\log g=4.5$ (dashed line) and 5.0 (dot-dashed line). Lower
  panel: Same as before but for GJ-866 which is an older field dwarf
  having the same spectral type than BL-29 (M6).}
  \label{fitNa}
\end{figure}

Knowing the effective temperature of the candidates, we then compare
the gravity sensitive Na~I doublet (8183{\AA}, 8195{\AA}) profile to
synthetic spectra of similar $T_{eff}$ given by Allard et al. (2001)
for various $\log g$. The best fit allows us to estimate the surface
gravity of the objects with an accuracy of 0.5 in $\log g$
corresponding to the grid step. Since the expected value of $\log g$
for $\sim100$ Myr \bds ($\log g\simeq 4.8$) is lower than the one for
older field dwarfs ($\log g\simeq 5.3$), we can use this estimate
as a membership criterion. Figure~\ref{fitNa} shows the comparison of
the Na~I doublet profile of a very low mass candidate (BL-29) and a
similar spectral type (M6) standard field dwarf GJ-866 with two
synthetic spectra of similar $T_{eff}$ for $\log g=4.5$ and 5.0. The
older field object is better fitted by the spectrum of larger gravity
as expected, while the best fit of BL-29 profile is obtained for $\log
g=4.5$. This indicates that this object is likely a cluster member. We
did this analysis for all the candidates for which we have a spectrum
with a good enough signal to noise ratio to estimate their surface
gravity. When possible we also measure the equivalent width of the
$H_{\alpha}$ emission line (6563{\AA}), indicative of chromospheric
activity. All the results are indicated in Table~\ref{bl_spectro}.

All the candidates observed spectroscopically are late M dwarfs which
confirms that the contamination by galaxies or red giants is indeed
very low. CFHT-BL-44 has an earlier PC3 spectral type and a hotter
effective temperature than the other candidates of similar
$I$-magnitude. Moreover its surface gravity is too large to be a
cluster member. It is most likely a foreground field dwarf. We
suspect CFHT-BL-28 is also a field dwarf for the same reasons (slighty
earlier PC3 spectral type and hotter temperature) even though the
analysis of the Na profile did not provide a definitive result.
CFHT-BL-34 and -41 have a later spectral type and a lower temperature
than expected from their $I$-magnitude and both $I-z$ and $I-K$
colours redder than the NEXTGEN isochrone. Their surface gravity is
low and consistent with cluster membership and we therefore believe
these two objects are cluster binaries. Note that three other objects
(CFHT-BL-52, -53, -54) are also suspected to be cluster binaries from
their location above the isochrone in the infrared CMD but we could
not get any spectra as they are too faint.

Between $I=18.5$ and 20, three objects out of 34 photometric
candidates have been identified as non-members, either because of
their infrared colour (CFHT-BL-47) or because of their spectroscopic
properties (CFHT-BL-28 and 44). Eight objects do not have any
follow-up observation, neither $K$ photometry nor spectroscopy, and we
cannot conclude on their nature. This yields a contamination level
between 10 and 30\%, in agreement with the statistical estimate given
above. We can refine this analysis if we note that all the objects
without any follow-up have about the same $I-z\simeq 0.74$ and form a
small group in the CMD (see Fig.~\ref{cmdlong}), slighty bluer than
the candidates that are confirmed spectroscopically. Three other
objects that have infrared follow-up lie also in this group. All of
them are bluer than the 100 Myr (and sometimes also than the 150 Myr)
isochrone in $I-K$, and one is a non-member (CFHT-BL-28) based on its
spectroscopy. Therefore the contamination level in this subgroup is
probably high, larger than 50\%, which gives a total contamination
level around 20-30\% for $I=18.5\le I\le20$.

We can use the same argument to estimate the number of contaminants in
the magnitude range $I=21-22.2$ corresponding to our sample
completeness limit. We do not have any $K$-band photometry for nine
objects that are amongst the bluest ($I-z\simeq1.0$) candidates in
this region of the ($I,I-z$) CMD. Considering that about half of the
objects with $I-z=1.05-1.10$ have been identified as non-members
(CFHT-BL-73, -79, -84, -95, -96) and that there is no contaminant
amongst the redder objects, we expect that about five or more of these
blue candidates are field dwarfs.

As for the short exposures, we did the full analysis for both
selections based on the 100 Myr and 150 Myr isochrones
respectively. The number of estimated contaminants and of probable
cluster members per magnitude bin in the long-exposures images is
summarized in Table~\ref{lf_bd} for each age.

\begin{table*}[htbp]
  \centering
  \caption{Number of very low mass star and brown dwarf candidates,
  estimated contaminants and probable cluster members after analysis
  of the follow-up observations (infrared photometry + optical
  spectroscopy) for both selections (100 Myr and 150 Myr).}  
  \input{6308tab5.tex}
  \label{lf_bd}
\end{table*}

\subsection{Spatial distribution}

Now that we have identified the cluster member candidates with
$0.03\msun\le m \le 0.6\msun$ in the survey, we need to study their
spatial distribution. It is necessary to check whether mass
segregation has occured in the cluster in order to take it into
account when determining the mass function.

\begin{figure}[htbp]
  \includegraphics[width=0.9\hsize]{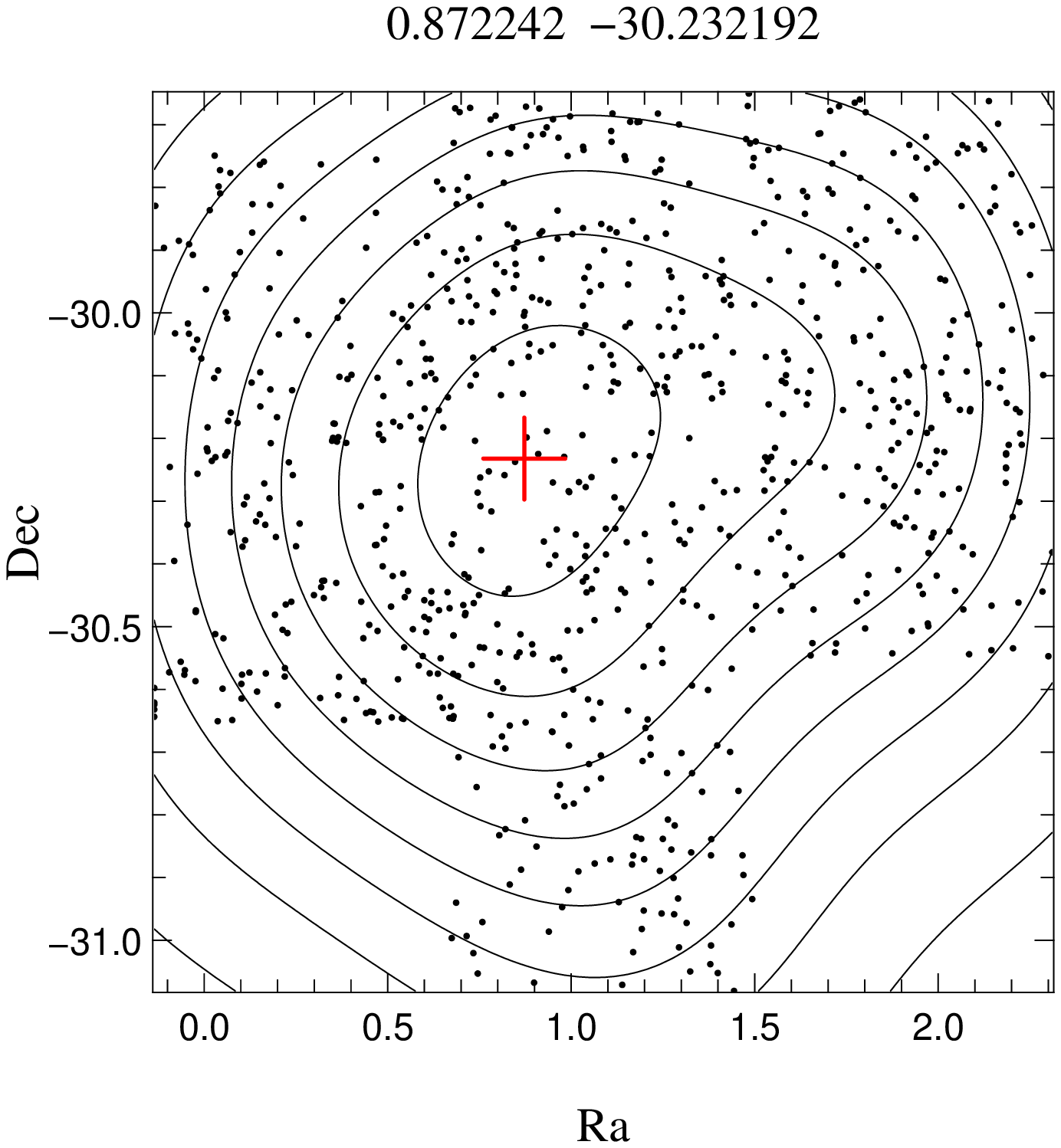}
  \caption{Spatial distribution of all the cluster member candidates
  detected in the short exposures and selected from the 150 Myr
  isochrone. Contours of isodensity are drawn. The cross represents
  the peak of density and is taken as the cluster center whose
  coordinates are (0.87,-30.2) in degrees.}
  \label{distrib}
\end{figure}

Figure~\ref{distrib} shows the spatial distribution of the low mass
star candidates selected in the short exposures when assuming a
cluster age of 150 Myr. Using the kernel method described by Silverman
(1986) we estimated the isodensity contours. The peak of the
distribution allows us to determine the coordinates of the cluster
center: $\alpha_{0}=0^{\rm h}04^{\rm m}1.2^{\rm s}$,
$\delta_{0}=-30\degr 12'$. This is about $0.3\degr$ away from the
nominal center given by Lyng\aa\, (1987) but close to the recent
estimate given by Kharchenko et al. (2005) who used the maximum of the
projected stellar density as the definition of the cluster center.

\begin{figure}[htbp]
  \includegraphics[width=0.9\hsize]{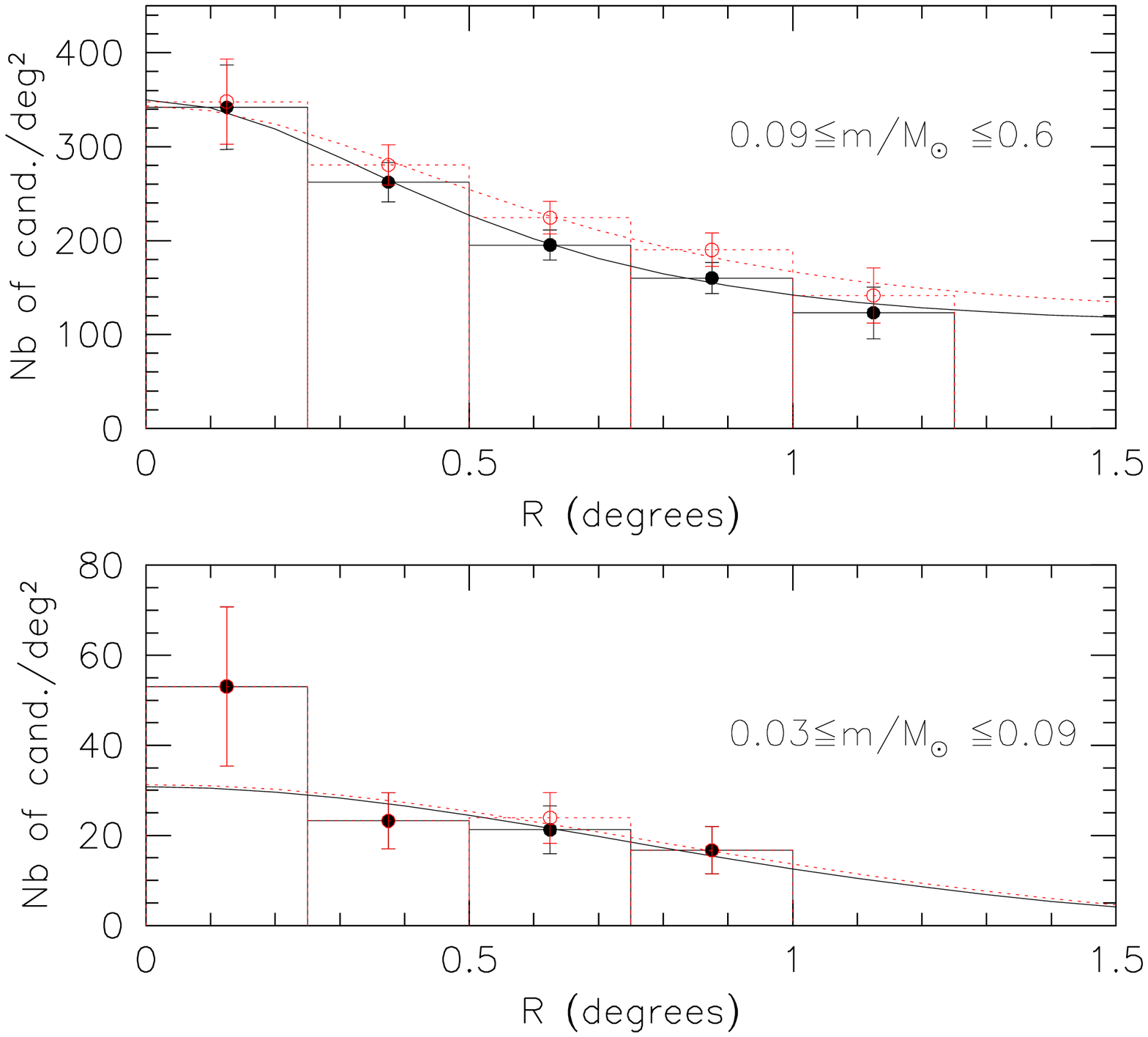}
  \caption{Radial distribution of the low mass star candidates (upper
  panel) and the VLM \& BD probable cluster members (lower panel). The
  cluster center has been estimated using the isodensity contours
  (Fig.~\ref{distrib}). In both cases, the filled dots and solid
  histogram represent the 100 Myr sample while the empty dots and
  dashed line histogram correspond to the 150 Myr sample. The best
  $\chi^2$ fit of a King profile is given for each histogram.}
  \label{hist}
\end{figure}

We then calculate the number of candidates per square degree located
within each annulus $R, R+dR$ centered on the new center ($\alpha_{0},
\delta_{0}$). The radial profiles obtained for the low mass star
candidates ($0.09\msun\le m\le0.6\msun$, short exposures) and for the
very low mass star and brown dwarf candidates ($0.03\msun\le
m\le0.09\msun$, long exposures) are shown Figure~\ref{hist}.

We assume that the distribution of the low mass sample contaminants is
uniform and we note their spatial density $n_{cont}$. The low mass
star distribution is then fitted by a King profile (King 1962) plus
$n_{cont}$
  \begin{equation}
    \label{eq:1}
    n(x) = k\,\left[\frac{1}{\sqrt{1+x}} -
      \frac{1}{\sqrt{1+x_{t}}}\right]^2 + n_{cont}
  \end{equation}
where $k$ is a normalisation constant, $x=(r/r_{c})^2$ and
$x_{t}=(r_{t}/r_{c})^2$. The distance from the cluster center
($\alpha_{0}, \delta_{0}$) is $r$, $r_{c}$ the core radius and $r_{t}$
the tidal radius.

According to Pinfield et al. (1998), the tidal radius $r_{t}$ of a
cluster close to the Sun in a circular orbit is given by
\begin{equation}
  r_{t} = \left( \frac{GM_{c}}{2(A-B)^2} \right)^{1/3} 
        = 1.46 \, M_{c}^{1/3}
\end{equation}
where $M_{c}$ is the cluster mass and $A$ and $B$ are the Oort
constants. Blanco~1 contains about 180 F, G and K stars (Abraham de
Epstein \& Epstein 1985, Jeffries \& James 1999, Pillitteri et
al. 2003), i.e. about half as many as the Pleiades. If we assume that
both clusters have the same mass distribution, then we expect Blanco~1
is about half as massive as the Pleiades and $r_{t}({\rm
Blanco1})=0.794\times r_{t}(\rm Pleiades)$. The Pleiades tidal radius
being 13.1 pc (Pinfield et al. 1998), we derive $r_{t}({\rm Blanco1})=
10.4$ pc or $2.3\degr$ for a distance of 260 pc. 

Using this value, the best fit minimizing $\chi^2$ gives
$n_{cont}=112\pm10$, $k=430\pm40$ per square degrees and
$r_{c}=0.61^{+0.16}_{-0.11}$ degrees, i.e $2.8^{+0.7}_{-0.5}$ pc, for
the 100 Myr sample. For 150 Myr, we find $n_{cont}=125\pm10$,
$k=479\pm43$ and $r_c=0.79^{+0.15}_{-0.16}$ degrees or $3.6\pm0.7$
pc. Note that the values obtained for $n_{cont}$ are consistent with
the number of contaminants given in Table~\ref{contam} divided by the
surveyed area (2.3 sq.deg.). We find $n_{cont}=92\pm13$ sq.deg$^{-1}$
for the 100 Myr selection and $n_{cont}=127\pm17$ sq.deg$^{-1}$ for
the 150 Myr one. This validates {\it a posteriori} our statistical
contamination estimate. We then compute the total number $N$ of
cluster members in the mass range 0.09-0.6$\msun$ by integrating
equation~\ref{eq:1} for both ages and by adopting the mean value. We
find $N=380\pm50$ stars ($N_{\rm 100 Myr}=343^{+30}_{-35}$ and $N_{\rm
150 Myr}=416\pm37$). According to table\ref{contam}, about $255\pm60$
of them are present in our survey, i.e. $\sim67\%$.

The radial distributions of the 49 (for the 100 Myr selection) and 51
(150 Myr selection) probable VLM \& BD members remaining after
analysis of the follow-up observations are also fitted by a King
profile (see Fig.~\ref{hist}). Note that no constant is added to the
King distributions as the samples used to plot the histograms have
been corrected from the contamination (the candidates without any NIR
photometry or optical spectroscopy are not taken into account
here). The fit parameters corresponding to the 100 Myr selection
are $r_{c}=1.56^{+0.16}_{-0.15}$ degrees ($7.1\pm0.7$ pc) and
$k=160\pm23$ sq.deg.$^{-1}$. For the 150 Myr sample, we obtain
$r_{c}=1.70^{+0.16}_{-0.14}$ degrees and $k=190\pm27$
sq.deg.$^{-1}$. The integration of both fits yields $104\pm15$ and
$112\pm15$ objects respectively, indicating that we identified about
47\% of them in the CFH12K images. Among the 17 candidates without any
follow-up observations, we estimate that $\sim9$ are probable members
(see previous section). Assuming that they follow the same radial
distribution, this yields a total number of $125\pm20$ VLM \& BDs in
the cluster.

About 47\% of the Blanco~1 very low mass stars and brown
dwarfs have been covered by the survey whereas $\sim$67\% of the low
mass star members are present in our study. This difference has to be
taken into account when estimating the cluster mass function. It
also suggests that mass segregation due to two body interactions has
already occured and that Blanco~1 is dynamically relaxed. The
relaxation time $t_{r}$ is
\begin{equation}
t_{r}=0.8 \frac{\sqrt{N_c}\, r_h^{3/2}}{\sqrt{\bar{m_c}}\,(\log N_c -
  0.3)} \,\, {\rm Myr}
  \label{tr}
\end{equation}
(Spitzer 1940), where $N_c$ is the total number of cluster members,
$r_h$ the half mass radius in pc and $\bar{m_c}$ the average stellar
mass. The crossing time $t_{cr}$ is then given by King (1980) by
\begin{equation}
t_{cr}= t_{r} \,\frac{31\,\ln \frac{N_c}{2}}{N_c}.
  \label{tcr}
\end{equation}
As $N_c({\rm Blanco~1})<N_c(\rm{Pleiades})$, $r_{h}({\rm
Blanco~1})\lesssim3$ pc while $r_{h}({\rm Pleiades})=3.66$ pc
(Pinfield et al. 1998), and $\bar{m_c}({\rm Blanco~1})\sim
\bar{m_c}({\rm Pleiades})$, we find that these two dynamical times are
smaller for Blanco~1 than for the Pleiades. In particular, this yields
$t_r(\rm{Blanco~1})\le 90$ Myr which is smaller than its age around
100-150 Myr. This suggests that Blanco~1 is indeed dynamically
relaxed. Moreover, when complete equipartition of energy is reached,
the cluster core radius varies with stellar mass as $m^{-0.5}$, and we
find that this is roughly the case here. For both ages, we find
$r_{c}=(1.6\pm0.2)\, \bar{m}^{-0.5}$ in pc where $\bar{m}$ is the mean
mass of the sample ($\sim0.3$ for low mass stars and $0.05\msun$ for
VLMs \& BDs).

\subsection{Mass function}

The number of probable \bl members per $I$-magnitude bin is given in
Tables~\ref{contam} and ~\ref{lf_bd}, last column. In order to
estimate the cluster {\it system} mass function -- multiple systems
are not resolved at the distance of the cluster -- we first convert
the magnitude bin to a mass bin using the 100 Myr or 150 Myr
mass-magnitude relationship from the NEXTGEN models (Baraffe et
al. 1998) for $I\le 20.0$ and DUSTY models (Chabrier et al. 2000) for
$I\ge 21.0$. The obtained mass ranges are indicated in
Table~\ref{contam} and ~\ref{lf_bd}, second column. Then we divide the
number of probable members by a correction factor corresponding to the
percentage of low mass stars and brown dwarfs covered by our survey
(i.e. respectively 0.65 and 0.48).

\begin{figure}[htbp]
  \includegraphics[width=0.9\hsize]{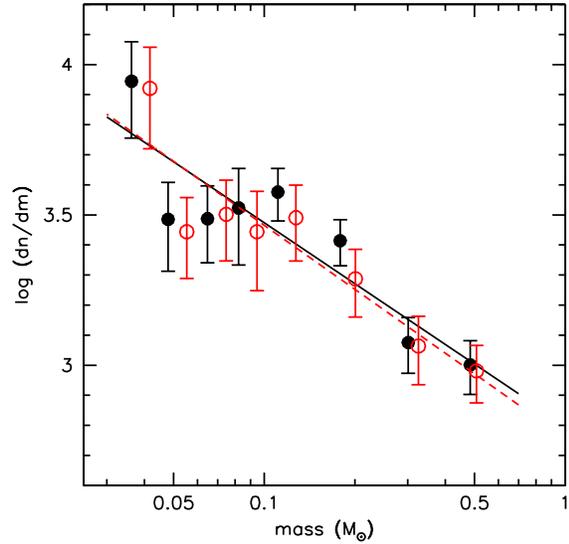}
  \caption{The Blanco~1 system mass function across the
    stellar/substellar boundary. Note that all the data points are
    derived from the same survey, using short exposures for the
    stellar domain and long exposures for the substellar regime. This
    provides a consistent determination of the slope of the cluster's
    mass function in the mass range from 0.030 to $0.6\msun$. The data
    points are fitted by a power law $dn/dm \propto m^{-\alpha}$ with
    an index $\alpha=0.67\pm 0.14$ for the 100 Myr sample (solid
    line) and $\alpha=0.71\pm 0.13$ for the 150 Myr sample (dashed
    line).}
  \label{imf}
\end{figure}

The total number of cluster systems per unit mass is thus derived over
a continuous mass range from 0.03 to 0.60$\msun$ and is shown in
Fig.~\ref{imf}. In this mass domain, the mass function can be fitted
by a single power-law $dn/dm \propto m^{-\alpha}$. A linear regression
through the data points yields $\alpha=0.67\pm 0.14$ for the 100 Myr
sample and $\alpha=0.71\pm 0.13$ for the 150 Myr sample where the
uncertainty is the $1\sigma$ fit error. (In this representation,
Salpeter's IMF corresponds to $\alpha=2.35$.)

\begin{figure}[htbp]
  \includegraphics[width=0.9\hsize]{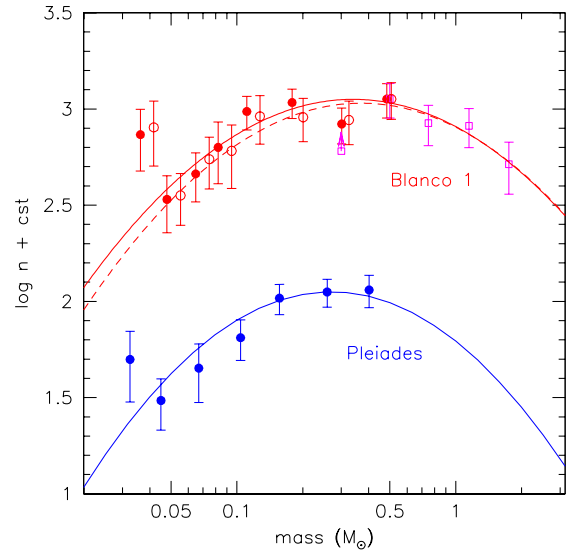}
  \caption{The mass function of Blanco~1, from low mass brown dwarfs
  to the most massive stars, fitted by a log-normal distribution
  (solid line and dashed line for the 100 Myr and 150 Myr sample
  respectively). The large filled (100 Myr sample) and open (150 Myr
  sample) dots are our data points completed in the more massive star
  domain by Pillitteri et al. (2003) data (open squares). The Pleiades
  mass function (data points and log-normal fit from Moraux et
  al. 2003) is shown for comparison (see Section 4).}
  \label{imflog}
\end{figure}

At larger masses, we completed our study with data from the
literature. Pillitteri et al. (2003) established a list of reliable
cluster members based on proper motion measurements. Their sample
contains objects with mass ranging from 3$\msun$ down to
$\sim0.2\msun$ but is not complete below $\sim0.4\msun$. Using the
mass estimate given in their paper (table A1)\footnote{Note that the
authors used an age of 100 Myr to estimate masses but using 150 Myr
instead would not change the results in this mass range. This is not
the case however at lower mass where the 100 Myr and 150 Myr
mass-magnitude relationships differ.}, we compute the number of
objects per unit mass and multiply it by a constant, so that the
number of stars in the range 0.4-0.6$\msun$ is the same as the one we
derive from our survey. On the whole mass range, from about 30 Jupiter
masses to 3$\msun$, the Blanco~1 system mass function corrected for
contamination is reasonably well fitted by a Scalo-like log-normal
distribution
\begin{displaymath}
    \xi_{\rm L}(m) = \frac{dn}{d\log m} \propto \exp\left[ -
      \frac{(\log m - \log m_{0})^2}{2\sigma^{2}}\right]
\end{displaymath}
where $m_{0}$ corresponds to the peak of the distribution and $\sigma$
represents its width (see Fig.~\ref{imflog}). A $\chi^{2}$ fit gives
$m_{0}=0.34\pm0.05\msun$, $\sigma=0.58\pm0.06$ for the 100 Myr sample,
and $m_{0}=0.38\pm0.05\msun$, $\sigma=0.58\pm0.06$ for the 150 Myr sample.

By integration of the log-normal fit, we find a total number of brown
dwarf systems ($m=0.01-0.072 M_{\odot}$) in the cluster of about 95
(85), for $\sim660$ (690) stars in the mass range $0.072-3
M_{\odot}$\footnote{Note that the upper mass limit is imposed by the
highest mass cluster member with $m\sim 3\msun$ ($M_{(bol)}=-0.74$;
Westerlund et al. 1988).}, which yields a ratio of about 15\% (12\%)
for the 100 Myr (resp. 150 Myr) sample. The cluster total mass is
$\sim 410\msun$ for an age of 100 Myr ($\sim 450\msun$ for 150 Myr)
with a substellar contribution corresponding to only $\sim1\%$. The
stellar mass of the Pleiades is $735\msun$ (Pinfield et al. 1998),
which confirms {\it a posteriori} our initial assumption of Blanco~1
being about half as massive as the Pleiades. From equations~\ref{tr}
and~\ref{tcr}, we find $t_{r}\lesssim60$ Myr and $t_{cr}\lesssim15$
Myr, which indicates that the cluster is dynamically relaxed indeed.

\section{Discussion}

Blanco~1 has an age similar to, or a little older (Panagi et al. 1997)
than, the Pleiades but it is less dense (about 30 stars/pc$^2$ against
$\sim65$ stars/pc$^2$), about half as massive and has a different
abundance pattern (Ford, Jeffries \& Smalley 2005). Comparing the mass
functions of these two clusters thus allows us to test the dependence
of the IMF on environmental conditions, especially as they have been
obtained from similar data sets and analysed in the same way. Both
mass functions are plotted on Figure~\ref{imflog}. The Pleiades MF is
from Moraux et al. (2003), their data points are between $0.03\msun$
and $0.5\msun$ and the log-normal fit goes from $0.03\msun$ to
$3\msun$.

As seen from both Figures~\ref{imf} and~\ref{imflog}, the
Blanco~1 mass function does not depend much on its exact age. The
results are similar within the uncertainties. In the following we will
do the comparison with the Pleiades MF for the 100 Myr Blanco~1 MF
only but the conclusions remain valid for 150 Myr.

A quick look at Figure~\ref{imflog} is sufficient to note that the
overall shapes of the Pleiades and Blanco~1 MF are very similar. Both
mass functions can be fitted by a log-normal distribution with a
characteristic {\it system} mass around $m_{0}\simeq0.3 M_{\odot}$ and
$\sigma\simeq0.5$. It is interesting to add that this is also the case
for a large number of star forming regions and other open clusters, as
well as for the system galactic disk mass function down to $0.1\msun$
at least (see Chabrier 2003). This suggests either that the IMF does
not depend much on environmental conditions or that star formation
occurs only for a narrow range of local parameters as discussed in
section~\ref{cond}.

\subsection{The M7/8 gap}
\label{M78}
By looking at the Blanco~1 and Pleiades data points, we also notice
that for both clusters the point around $0.035\msun$ is much above the
log-normal fit. This cannot only be due to an underestimate of
the contamination in this mass range as it is taken into account in
the error bar. It is as if the IMF rises again below
$0.04\msun$. Similarly, Muench et al. (2002) found a significant
secondary peak close to the deuterium limit in the Trapezium
IMF. However, the fact that this feature does not occur around the
same mass but the same spectral type (later than M8) brought Dobbie et
al. (2002) to suggest that this is not real. They argue instead that
this structure reflects a sharp local drop in the luminosity-mass
(L-M) relationship due to the onset of dust formation in the
atmosphere around $T_{eff}=2700$~K. A change in the slope of the L-M
relationship implies that the mass of objects with spectral type later
than M7-M8 may be significantly underestimated by the current NextGen
and Dusty models. Thus, the number of objects in the lower mass bin
may be overestimated, resulting in an artificial secondary peak in the
IMF.

\begin{figure}[htbp]
  \includegraphics[width=0.9\hsize]{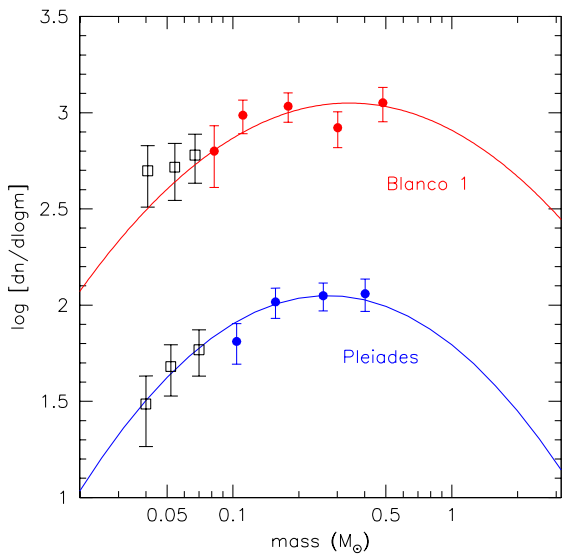}
  \caption{The Blanco~1 and Pleiades mass function obtained after
    applying the empirical magnitude-mass relation given by Dobbie et
    al. (2002). The open squares are the data points which have been
    modified while the filled circles are the data points which did not
    change. The log-normal fit for each cluster is shown as a
    solid line.}
  \label{mfM78}
\end{figure}

By applying the empirical magnitude-mass relation given by Dobbie et
al. (2002; their Fig.3) to the Blanco~1 and Pleiades luminosity
function, we find that the lowest mass point goes down for both MFs
whereas higher mass points go slightly up (see Fig.~\ref{mfM78}),
following nicely the log-normal fit. Even if this dust effect has
still to be investigated, we consider the log-normal distribution is a
good approximation to the observed MF, including in the substellar
domain down to 30 Jupiter masses.

\subsection{Dependency of the IMF on environmental conditions}
\label{cond}

With an age of about 100 Myr, the Blanco~1 and Pleiades clusters are
already dynamically evolved in the sense that they are well relaxed
and that mass segregation has occured. However they are still young
enough as not to have lost many objects. Adams et al. (2002) estimated
that only $\sim10$\% of the primordial low mass star population has
been lost in the Pleiades, which has been confirmed by other studies
combining numerical simulations with observations (e.g. Moraux, Kroupa
\& Bouvier, 2004). Therefore we can assume that the present day mass
functions discussed here represent well the IMF for both clusters.

We have already mentioned that, despite the different local parameters
(stellar density, abundance pattern), these 2 IMFs are similar within
the uncertainties over 2 decades in mass. This seems also to be the
case for other young open clusters such as Alpha-Per ($\sim80$ Myr,
Barrado et al. 2002), IC2391 ($\sim50$ Myr, Barrado et al. 2004a),
NGC2547 ($\sim30$ Myr, Jeffries et al. 2004) or IC4665 ($\sim30$ Myr,
De Wit et al. 2006). One could argue that only one ``type'' of
clusters, rich and compact enough, can survive several tens of
Myr. Therefore they cannot be very different initially and it is not
very surprising to find a similar IMF. However most of the studied
star forming regions ($\lambda$-Ori, Barrado et al. 2004b; IC348,
Luhman et al. 2003a; Trapezium, Muench et al. 2002; $\sigma$-Ori,
B\'ejar et al. 2004) also present similar IMF's within the
uncertainties. Again, it is true that they are all fairly rich but
even in Taurus, a much sparser region, recent results from Guieu et
al. (2006) seem to indicate that the IMF is not very different (but
see also Luhman et al. 2003b and Briceno et al. 2002 for a differing
view). Probably even more remarkable is the similarity with the
galactic disk mass function (Chabrier 2003) -- although uncertainties
remain in the substellar domain -- as field stars come from a large
variety of environments, from small groups with $N_{star}<10$ to rich
clusters (e.g. Allen et al. 2006). All these results suggest that the
IMF, from the substellar regime to the high mass domain, is similar in
all studied regions. The fact that the IMF does not seem to depend
much on the environment brings strong constraints on star formation
theories. To date, two main models (turbulent fragmentation and
accretion/ejection) are proposed to explain the origin of the IMF;
they are discussed below.

\subsubsection{Turbulent fragmentation}

Turbulent fragmentation can generate a distribution of density
structures in molecular clouds due to supersonic shocks that compress
the gas. Multiple compressions result in the formation of sheets and
then filaments, whose density and width are due to the MHD shock
conditions. High velocity shocks produce high density but thin
filaments resulting in low-mass clumps, whereas low velocity shocks
produce low-density but large clumps which account for high mass
objects. Using the power spectrum of velocities from numerical
simulations of turbulence, Padoan \& Nordlund (2002) derive a clump
mass distribution resulting in a IMF slope which closely matches
Salpeter's. At lower masses however, only a fraction of these clumps
are sufficiently dense to be unstable which results in a flattening
and a log-normal shape in the substellar regime.

A major success of this model is that it is able to reproduce directly
the observed prestellar core distribution (Motte \& Andr\'e 2001)
which resembles effectively the stellar IMF at least down to
$0.1\msun$. However, how the clump spectrum translates into the
stellar mass distribution is not fully understood. In particular,
it is very likely that the higher mass clumps will sub-fragment into
smaller clumps and form several stars. Moreover, SPH simulations find
that most of the small cores are unbound and do not collapse to form
stars (Klessen et al. 2005).

An important result of the turbulent fragmentation model is that the
shape of the IMF depends on the turbulence via the Mach number or the
power spectrum (Ballesteros-Paredes et al. 2006, Goodwin et al. 2006)
but also on the molecular cloud density (see Padoan \& Nordlund 2002,
their fig.~2). This is easy to understand using simple
considerations. Typically, a denser cloud forms denser but smaller
clumps yielding a larger number of low mass cores. Conversely, at
lower density, the mass function peaks at higher mass. This result
concerns the core mass function and it is not completely clear yet how
this will affect the stellar IMF. However it is reasonable to think
that the trend will be the same, which is in contradiction with the
observed invariance of the IMF.

\subsubsection{Accretion/ejection paradigm}

Another model which has been developed to explain the origin of the
IMF origin of the IMF gives less importance to turbulence and more to
gravity. According to Bate \& Bonnell (2005), the role for turbulence
is to generate structure in molecular clouds which provides the seeds
for gravitational fragmentation to occur down to the opacity
limit. Stellar masses are then set by a combination of accretion and
dynamical ejections which terminate the accretion. In their model
based on the results of hydrodynamical SPH calculations, all the
objects begin with a mass corresponding to the opacity limit ($\sim
0.003\msun$) and grow by accretion until they are ejected from the
dense core by dynamical interactions in unstable multiple systems. The
objects that end up as brown dwarfs stop accreting before they reach
stellar masses because their are ejected soon after their
formation. In contrast the higher mass stars are the objects that
remain in the dense gas and accrete for longer. The accretion rates of
individual objects are drawn from a lognormal distribution and the
dynamical ejection of protostars from a multiple system is described
using a characteristic time-scale. Taking values for these parameters
from the hydrodynamical simulations, the model is able to reproduce a
reasonable IMF in the tellar domain but produces more brown dwarfs
than are actually observed.


In the accretion/ejection model the characteristic (median) mass of
the IMF is approximatively given by the product of the typical
accretion rate and the typical time-scale for dynamical ejection. Bate
\& Bonnell (2005) and Bate (2005) found that that the characteristic
mass varies linearly with the mean thermal Jeans mass $M_{Jeans}$ and
depends therefore on both density $\rho$ and temperature $T$. As
$M_{Jeans}$ is smaller at higher density, the mass function is
expected to peak at smaller mass. Indeed in a denser cloud dynamical
ejections are more likely to occur, yielding the formation of a larger
number of low mass stars. Variations of the IMF in different
environments should then occur in the location of the peak and in the
substellar regime, which is not supported by observations.

\subsubsection{Thermal physics}

The two main models developed to explain the origin of the IMF predict a
dependence of the IMF on the environment, in particular on the density
and temperature of the molecular cloud via the thermal Jeans mass
and/or turbulence.

Observational results however indicate that the IMF is similar in many
regions from the substellar regime up to the massive star domain. This
suggests that either the global properties of all the molecular clouds
in which cluster and field stars form are very similar -- which seems
very unlikely -- or that the physical process for fragmentation by
itself imposes specific local conditions. Larson (2005) has suggested
that thermal physics in molecular clouds may play this role as it can
yield a Jeans mass of order $\sim1\msun$ at the point of fragmentation
independently of the exact initial conditions. This results from how
the cooling rate changes with density. At lower densities the gas
cooling is dominated by atomic and molecular line emission whereas at
higher densities the gas is coupled to the dust and dust cooling
dominates. This results in a barotropic equation of state involving a
cooling term at low densities followed by a gently heating term once
dust cooling prevails. When used in numerical calculations, this
equation of state can set the characteristic mass scale for
fragmentation and produce a realistic IMF (Jappsen et al. 2005,
Bonnell et al. 2006a).

A prediction of thermal physics is that a different metallicity may
yield a different IMF as the cooling equation of state is modified. If
a molecular cloud is more metal rich, the cooling is more efficient
and the temperature is smaller. Therefore $M_{Jeans}$ is smaller and
the characteristic mass is shifted towards the lower mass. However,
the thermal physics invoked by Larson (2005) links density and
temperature.  Thus for a higher metallicity, the Jeans mass occurs at
lower $T$ but also lower $\rho$ and whether it should be larger or
smaller is not clear yet. It is therefore premature to give any trend
for the dependency of the IMF on metallicity but it could be a good
observational test to measure and compare the mass function of stellar
clusters with different metallicity.

\subsubsection{An hybrid model ?}

Bonnell et al. (2006b) have suggested an ``hybrid'' model to explain
the origin of the mass distribution. Turbulence is necessary to
generates the filamentary structure in the molecular clouds which
facilitates fragmentation. Thermal physics sets the mean Jeans mass
for gravitational fragmentation which corresponds to the
characteristic stellar mass of the IMF. The broad peak can then be
understood as being due to the dispersion in gas densities and
temperature at the point where fragmentation occurs. The higher mass
part is due to the continued competitive accretion in dense
cores while the lower mass IMF is ascribed
to fragmentation and then ejection. Some of the collapsing regions,
especially in filaments and circumstellar discs, sub-fragment due to
the increase of gas density. The lower mass clumps are ejected soon
after they form from their natal environment by dynamical
interactions, stop accreting and remain low mass objects (very low
mass stars and brown dwarfs).

This model seems to work at reproducing a realistic stellar IMF as the
location of the peak and the higher mass part do not vary with initial
conditions as shown by Bonnell et al. (2006a). However some issues
remain in the {\it substellar} domain. In particular, the lower mass
part of the IMF is still expected to depend on the initial density --
denser clouds enhance dynamical ejections and thus form more brown
dwarfs -- which is in contradiction with the current observations.


\section{Conclusion}

We performed a deep large-scale photometric survey of the young open
cluster Blanco~1 to study its low mass population, from about 30
Jupiter masses to $0.6\msun$. We selected cluster candidates on the
basis of their location in CMDs compared to theoretical isochrones and
we estimated the contamination using statistical arguments, infrared
photometry and optical low-resolution spectroscopy when possible. We
did this analysis for two cluster ages (100 and 150 Myr) and we find
similar results in both cases. We estimate that about 300 cluster
members have been covered by our survey, amongst which 30-40 may be
brown dwarfs. The study of the candidate radial distribution indicates
that this corresponds to $\sim 57\%$ of the cluster low mass
population.

It also suggests that mass segregation has already occured in the
cluster and we took it into account when estimating the mass
function. We find that a single power-law $dN/dM \propto M^{-\alpha}$
with $\alpha=0.69\pm0.15$ provides a good match to the cluster MF
accross the stellar/substellar boundary, in the $0.03-0.6\msun$ mass
range for both ages. When complementing our survey with literature
data from Pillitteri et al. (2003) in the higher mass range, we find
that the whole cluster mass function, from $0.03\msun$ to $3\msun$, is
well fitted by a log-normal distribution
\begin{displaymath}
    \xi_{\rm L}(m) = \frac{dn}{d\log m} \propto \exp\left[ -
      \frac{(\log m - \log m_{0})^2}{2\sigma^{2}}\right]
\end{displaymath}
with $m_{0}=0.36\pm0.07\msun$ and $\sigma=0.58\pm0.06$.

This result is very similar to the Pleiades MF given by Moraux et
al. (2003), the two clusters having about the same age but different
richness. Similar MF shapes are found for other young open clusters,
star forming regions and also for the galactic disc population. This
suggests that the IMF, from the substellar regime up to the higher
mass domain, is fairly insensitive to initial conditions and that
there is a characteristic system mass around $0.3\msun$ for star
formation. Theories developed to explain the origin of the mass
distribution, such as turbulent fragmentation (Padoan \& Nordlund
2002) or the accretion/ejection paradigm (Bate \& Bonnell 2005),
predict however a dependence of the IMF on density and temperature via
the thermal Jeans mass. This seems to be in contradiction with the
observational results unless thermal physics as invoked by Larson
(2005) sets the characteristic mass for fragmentation independently of
the initial conditions.

A clue to this problem may be found in studying the global properties
of collapsing clouds. The fact that all known embedded clusters, with
$N_{star}=10$ to more than 1000, have a stellar surface density which
varies by a factor of only a few (Allen et al. 2006) suggests that the
initial conditions may not be very different. Investigation of the
mass distribution of very low-mass pre-stellar cores should clarify
this issue (e.g. Li al. 2007, Nutter \& Ward-Thompson 2007).


\begin{acknowledgements}

The authors thank Isabelle Baraffe who calculated the NEXTGEN and
DUSTY models for the CFH12K filter set, as well as Sylvain Guieu for
computing the isodensity contours.

The authors wish to recognize and acknowledge the very significant
cultural role and reverence that the summit of Mauna Kea has always
had within the indigenous Hawaiian community. We are most fortunate
to have the opportunity to conduct observations from this mountain.

\end{acknowledgements}



\end{document}

%% file: 6308tab1.tex
\begin{tabular}{llcccc} 
\hline 
Field & RA(2000) & DEC(2000) \\
 & (h m s) & ($\degr$ $\arcmin$ $\arcsec$) \\
\hline
4  & $00:07:27.0$ & $-29:52:37$ \\
5  & $00:04:19.4$ & $-29:53:21$ \\
6  & $00:01:05.0$ & $-29:58:21$ \\
8  & $00:07:37.0$ & $-30:18:26$ \\
9  & $00:04:19.4$ & $-30:24:38$ \\
10 & $00:01:05.0$ & $-30:24:40$ \\
11 & $00:04:19.4$ & $-30:50:29$ \\
\hline 
\end{tabular}

%% file: 6308tab2.tex
\begin{tabular}{lllccc}
\hline
Assumed & $I$-magnitude & mass bin & Nb of selected & estimated nb of & nb of probable\\
age & bin       &  ($\msun$)  & objects & contaminants & members \\
\hline
& 14.0-15.5 & 0.60-0.39 & 120 & $52.42\pm8.25$  & $67.58\pm13.71$ \\
100 Myr & 15.5-16.5 & 0.39-0.23 & 120 & $58.22\pm7.02$  & $61.78\pm13.01$ \\
& 16.5-17.5 & 0.23-0.14 & 138 & $58.88\pm7.17$  & $79.12\pm13.76$ \\
& 17.5-18.5 & 0.14-0.09 & 101 & $41.58\pm6.10$  & $59.42\pm11.76$ \\
\hline
& 14.0-15.5 & 0.63-0.41 & 123 & $56.62\pm9.13$  & $66.38\pm14.37$ \\
150 Myr & 15.5-16.5 & 0.41-0.26 & 133 & $74.78\pm9.43$  & $58.22\pm14.90$ \\
& 16.5-17.5 & 0.26-0.16 & 149 & $85.94\pm10.34$ & $63.06\pm16.00$ \\
& 17.5-18.5 & 0.16-0.10 & 129 & $75.19\pm10.13$ & $53.81\pm15.22$ \\
\hline
\end{tabular}

%% file: 6308tab3.tex
\begin{tabular}{rcccrrl||rcccrr}
\hline
CFHT & $I$ & $I-z$ & $I-K_{s}$ & $RA_{J2000}$ & $DEC_{J2000}$ && 
CFHT & $I$ & $I-z$ & $I-K_{s}$ & $RA_{J2000}$ & $DEC_{J2000}$ \\
-BL- &&&& (h m s) & (d m s) && 
-BL- &&&& (h m s) & (d m s) \\
\hline
\it1&\it17.71&\it0.59&\it2.45&\it00 07 21.37&\it-30 15 06.37&&  63 & 20.64 & 1.08 & 3.90 & 00 06 49.32 & -29 40 40.97 \\ 
  2 & 17.81 & 0.63 & 2.70 & 00 01 14.56 & -29 46 06.34 &&  64 & 20.66 & 1.06 & 3.75 & 00 03 03.44 & -30 18 00.87 \\ 
  3 & 17.83 & 0.64 & 2.58 & 23 59 40.99 & -30 01 57.33 &&  65 & 20.74 & 1.10 & 4.10 & 23 59 52.62 & -29 54 27.01 \\ 
  4 & 17.84 & 0.69 & 2.72 & 00 02 08.12 & -30 10 50.78 &&  66 & 20.85 & 1.16 & 4.03 & 00 04 08.08 & -29 44 25.21 \\ 
  5 & 17.88 & 0.71 & 2.72 & 00 00 03.64 & -29 50 10.45 &&  67 & 20.97 & 1.01 & 3.97 & 00 05 05.14 & -29 49 55.71 \\ 
\it6&\it18.01&\it0.62&\it2.60&\it00 00 31.01&\it-30 06 39.81&&\it68 &\it21.03&\it0.98&-&\it00 05 30.34&\it-29 51 04.89 \\ 
  7 & 18.01 & 0.72 & 2.70 & 00 00 21.80 & -30 10 31.64 &&  69 & 21.21 & 1.11 & 3.97 & 00 01 20.12 & -30 02 30.66 \\ 
\it8&\it18.03&\it0.62&-&\it00 01 04.93&\it-29 44 59.60 &&  70 & 21.24 & 1.16 & 4.07 & 00 05 45.78 & -30 03 46.51 \\ 
  9 & 18.08 & 0.70 & 2.69 & 00 03 32.17 & -30 04 11.79 &&  71 & 21.24 & 1.06 & 3.80 & 00 04 30.77 & -29 56 16.09 \\ 
\it10&\it18.09&\it0.63&\it2.51&\it00 07 24.30&\it-30 18 29.32&&  72 & 21.28 & 1.19 & 4.07 & 00 06 08.95 & -29 44 25.25 \\ 
 11 & 18.19 & 0.74 & 2.80 & 00 02 12.96 & -30 01 20.53 &&  73 & 21.30 & 1.05 & 3.33 & 00 05 48.96 & -30 34 24.82 \\ 
 12 & 18.19 & 0.66 & 2.57 & 00 01 00.07 & -30 22 18.97 &&  74 & 21.37 & 1.19 & 4.26 & 00 03 43.73 & -30 04 01.75 \\ 
 13 & 18.23 & 0.71 & 2.84 & 00 07 06.85 & -30 27 34.52 &&  75 & 21.41 & 1.00 &  -   & 00 06 29.24 & -30 20 34.48 \\ 
 14 & 18.28 & 0.71 & 2.84 & 00 07 45.97 & -30 09 38.28 &&  76 & 21.42 & 1.11 & 4.15 & 00 04 58.71 & -30 53 28.03 \\ 
\it15&\it18.31&\it0.66&-&\it00 03 59.22&\it-29 41 11.96&&  77 & 21.66 & 1.06 & 3.82 & 00 05 17.81 & -31 02 58.90 \\ 
 16 & 18.33 & 0.73 & 2.87 & 00 01 28.46 & -30 06 06.56 &&  78 & 21.66 & 1.25 & 4.49 & 00 03 17.74 & -30 11 39.25 \\ 
\it17&\it18.34&\it0.65&-&\it00 00 01.76&\it-30 12 57.25&&  79 & 21.69 & 1.09 & 3.73 & 00 06 01.79 & -29 43 22.97 \\ 
\it18&\it18.38&\it0.66&-&\it00 05 25.13&\it-30 03 12.53&&  80 & 21.69 & 1.05 &  -   & 00 00 21.98 & -30 38 44.20 \\ 
 19 & 18.39 & 0.69 & 2.84 & 00 07 50.61 & -30 05 10.52 &&  81 & 21.70 & 1.00 &  -   & 00 06 33.15 & -30 30 48.80 \\ 
 20 & 18.42 & 0.69 & 2.69 & 00 02 58.2  & -30 45 21.0  &&  82 & 21.73 & 1.16 & 4.47 & 00 04 41.97 & -30 04 32.97 \\ 
\it21&\it18.49&\it0.67&-&\it23 59 54.46&\it-30 28 36.01&&  83 & 21.86 & 1.29 & 4.51 & 00 00 34.70 & -30 02 51.41 \\ 
 22 & 18.50 & 0.78 & 2.92 & 00 00 39.95 & -30 20 15.68 &&  84 & 21.87 & 1.04 & 3.39 & 00 07 24.34 & -30 26 18.90 \\ 
 23 & 18.52 & 0.79 & 2.99 & 00 00 48.86 & -30 02 03.42 &&  85 & 21.88 & 1.15 & 4.39 & 00 03 06.58 & -30 29 53.85 \\ 
 24 & 18.54 & 0.74 & 2.97 & 00 07 50.59 & -30 05 09.04 &&  86 & 21.93 & 1.01 &  -   & 00 05 31.32 & -30 50 59.03 \\ 
 25 & 18.65 & 0.79 & 3.08 & 00 00 42.70 & -30 17 43.71 &&  87 & 22.01 & 1.06 & 4.23 & 00 04 48.59 & -30 39 41.56 \\ 
 26 & 18.68 & 0.73 & 2.89 & 00 04 54.92 & -29 46 33.21 &&  88 & 22.01 & 1.24 & 3.98 & 00 03 52.50 & -31 01 49.98 \\ 
 27 & 18.78 & 0.73 &  -   & 00 00 52.81 & -30 33 58.02 &&  89 & 22.01 & 1.28 & 4.51 & 23 59 54.05 & -30 20 17.47 \\ 
 28 & 18.78 & 0.74 & 2.82 & 23 59 55.40 & -30 02 33.54 &&  90 & 22.02 & 1.03 &  -   & 00 06 05.77 & -30 15 32.67 \\ 
 29 & 18.80 & 0.81 & 3.05 & 00 04 20.23 & -30 46 20.98 &&  91 & 22.02 & 1.18 & 4.50 & 00 05 46.37 & -30 07 33.57 \\ 
 30 & 18.80 & 0.74 &  -   & 00 08 27.35 & -29 43 54.06 &&  92 & 22.07 & 0.99 &  -   & 00 00 52.62 & -30 07 57.62 \\ 
 31 & 18.80 & 0.74 &  -   & 00 03 45.56 & -30 24 04.88 &&  93 & 22.11 & 1.13 & 4.72 & 00 04 57.71 & -30 14 02.02 \\ 
 32 & 18.86 & 0.75 & 2.86 & 00 01 19.18 & -29 54 06.13 &&  94 & 22.13 & 1.07 &  -   & 00 08 17.92 & -30 02 46.07 \\ 
 33 & 18.90 & 0.75 &  -   & 00 06 10.76 & -30 21 37.50 &&  95 & 22.13 & 1.08 & 3.14 & 00 03 45.10 & -29 50 48.97 \\ 
 34 & 18.96 & 0.94 & 3.46 & 00 01 48.68 & -30 38 07.24 &&  96 & 22.17 & 1.14 & 3.83 & 00 03 20.65 & -30 50 59.23 \\ 
\it35&\it18.98&\it0.75&-&\it00 03 40.61&\it-29 54 49.03&&  97 & 22.19 & 1.01 &  -   & 00 03 27.14 & -29 50 06.02 \\ 
 36 & 18.99 & 0.89 & 3.13 & 00 07 08.74 & -30 06 43.68 &&  98 & 22.27 & 1.06 &  -   & 00 00 59.48 & -30 38 44.06 \\ 
\it37&\it19.00&\it0.74&-&\it00 03 58.51&\it-30 41 22.09&&  99 & 22.31 & 1.00 &  -   & 00 04 03.17 & -30 57 19.45 \\ 
 38 & 19.01 & 0.82 & 3.12 & 00 05 13.03 & -30 27 35.65 && 100 & 22.33 & 1.00 &  -   & 00 05 52.29 & -29 44 10.00 \\ 
\it39&\it19.02&\it0.74&-&\it00 02 43.01&\it-30 34 29.55&& 101 & 22.34 & 1.00 &  -   & 00 00 39.25 & -30 12 19.97 \\ 
 40 & 19.03 & 0.85 & 3.28 & 00 07 15.92 & -30 07 27.24 && 102 & 22.35 & 1.32 & 4.02 & 00 01 53.20 & -30 12 05.35 \\ 
 41 & 19.04 & 0.99 & 3.48 & 00 03 23.55 & -29 55 18.25 && 103 & 22.37 & 1.13 & 3.51 & 00 03 50.45 & -30 07 37.40 \\ 
\it42&\it19.10 &\it0.76&-&\it00 03 15.44&\it-30 35 54.63 && 104 & 22.38 & 1.07 &  -   & 00 04 06.67 & -30 05 55.52 \\ 
 43 & 19.11 & 0.83 & 3.15 & 00 04 32.81 & -30 18 42.32 && 105 & 22.38 & 1.04 &  -   & 00 01 05.95 & -30 03 12.58 \\ 
 44 & 19.15 & 0.85 & 2.89 & 00 03 52.15 & -30 32 57.41 && 106 & 22.39 & 1.06 &  -   & 00 01 53.21 & -30 12 03.86 \\ 
 45 & 19.33 & 0.89 & 3.29 & 00 01 35.62 & -30 03 09.47 && 107 & 22.43 & 1.13 & 3.75 & 00 03 28.01 & -29 43 45.54 \\ 
 46 & 19.37 & 0.87 & 3.07 & 00 05 57.00 & -29 43 47.66 && 108 & 22.52 & 1.12 &  -   & 00 05 44.87 & -29 55 26.88 \\ 
\it47&\it19.44&\it0.80&\it2.79&\it00 04 19.02&\it-30 37 15.91 && 109 & 22.54 & 1.07 &  -   & 00 00 26.33 & -30 12 17.60 \\ 
 48 & 19.47 & 0.87 & 3.45 & 00 08 13.8  & -30 16 50.1  && 110 & 22.56 & 1.26 & 3.47 & 00 07 48.99 & -29 54 36.75 \\ 
 49 & 19.49 & 0.92 & 3.58 & 00 04 28.83 & -30 20 37.52 && 111 & 22.56 & 1.26 & 5.31 & 00 00 35.43 & -30 04 53.36 \\ 
 50 & 19.69 & 0.92 & 3.47 & 23 59 50.09 & -30 01 59.69 && 112 & 22.57 & 1.10 &  -   & 00 02 57.95 & -29 40 22.47 \\ 
 51 & 19.81 & 1.02 & 3.56 & 00 04 07.61 & -29 59 18.63 && 113 & 22.59 & 1.09 &  -   & 00 00 49.57 & -29 45 44.17 \\ 
 52 & 19.83 & 1.00 & 3.88 & 00 02 06.57 & -29 44 57.86 && 114 & 22.61 & 1.14 & 4.92 & 00 02 05.63 & -30 01 23.50 \\ 
 53 & 19.83 & 1.01 & 4.31 & 00 02 16.02 & -30 18 41.41 && 115 & 22.63 & 1.18 & 3.93 & 00 04 31.85 & -30 28 21.95 \\ 
 54 & 19.84 & 1.04 & 3.88 & 00 07 41.38 & -29 56 19.93 && 116 & 22.69 & 1.21 & 3.70 & 00 06 18.68 & -30 18 24.42 \\ 
 55 & 19.95 & 1.02 & 3.62 & 00 02 15.04 & -30 09 52.53 && 117 & 22.71 & 1.12 &  -   & 00 01 58.71 & -30 25 03.15 \\ 
 56 & 20.07 & 1.04 & 3.80 & 00 03 37.35 & -30 51 02.15 && 118 & 22.74 & 1.28 & 5.30 & 00 06 32.55 & -29 46 05.90 \\ 
 57 & 20.07 & 0.95 & 3.32 & 00 07 22.70 & -30 01 57.00 && 119 & 22.74 & 1.22 & 3.86 & 00 03 17.81 & -30 26 16.41 \\ 
 58 & 20.15 & 1.00 & 3.68 & 00 08 04.50 & -29 56 53.33 && 120 & 22.78 & 1.25 & 4.42 & 00 00 00.77 & -30 34 21.89 \\ 
 59 & 20.31 & 1.02 & 3.72 & 00 04 27.44 & -30 35 20.00 && 121 & 22.80 & 1.27 & 4.47 & 00 00 20.80 & -29 44 55.84 \\ 
 60 & 20.46 & 1.05 & 3.88 & 00 03 40.13 & -30 03 40.86 && 122 & 22.87 & 1.25 & 2.90 & 00 07 57.97 & -30 26 06.88 \\ 
 61 & 20.49 & 1.13 & 3.92 & 00 04 11.85 & -30 45 06.08 && 123 & 22.99 & 1.41 & 4.43 & 00 09 02.9  & -30 19 05.3  \\ 
 62 & 20.56 & 1.13 & 3.87 & 00 03 55.54 & -30 34 10.74 && 124 & 23.00 & 1.28 & 4.41 & 00 01 12.14 & -30 00 25.73 \\ 
\hline
\end{tabular}

%% file: 6308tab4.tex
\begin{tabular}{llllllllc}
\hline
    & &\multicolumn{2}{c}{SpT from}& $T_{eff}$ from & & $\log g$ from && \\
BL- & PC3 & PC3 & reference & synthetic & $H_{\alpha}$ &
    synthetic & Instrument & comments \\
    & & & spectra & spectra (K) & EW(\AA) & spectra &&  \\
\hline 
16 & 1.26/1.27 & M4.9 & M5 & 2900 &  -10.0& 4.8 & FORS2/Keck02   &  \\ 
22 & 1.33      & M5.3 & M5 & 2900 &       & 5.0 & FORS2          &  \\ 
24 & 1.48      & M6.2 & M6 & 2850 &       &     & Keck (low s/n) &  \\ 
25 & 1.33      & M5.3 & M5 & 2850 &       & 5.0 & FORS2          &  \\ 
28 & 1.30      & M5.1 & M5 & 2950 &       & 5.0 & FORS2          & non-member ? \\ 
29 & 1.41      & M5.8 & M6 & 2850 &       & 5.0 & FORS2          &  \\ 
34 & 1.71      & M7.4 & M7 & 2700 &       & 4.5 & FORS2          & binary? \\ 
36 & 1.40/1.40 & M5.7 & M6 & 2850 & -5.2  & 5.0 & FORS2/Keck     &  \\ 
38 & 1.49      & M6.3 & M6 & 2800 & -6.5  & 4.8 & Keck           &  \\ 
41 & 1.77/1.71 & M7.6 & M8 & 2650 &$>-2.0$& 4.5 & FORS2/Keck     & binary? \\ 
43 & 1.49      & M6.3 & M6 & 2800 &       & 4.8 & FORS2          &  \\ 
44 & 1.33/1.39 & M5.5 & M6 & 2850 & -5.5  & 5.3 & FORS2/Keck     & non-member \\ 
45 & 1.50/1.61 & M6.6 & M7 & 2750 & -8.0  & 4.8 & FORS2/Keck     &  \\ 
48 & 1.67      & M7.2 & M7 & 2750 &       & 5.0 & FORS2          &  \\ 
49 & 1.73/1.62 & M7.2 & M7 & 2700 & -2:   & 4.8 & FORS2/Keck     &  \\ 
50 & 1.73      & M7.5 & M7 & 2700 &       & 4.8 & FORS2          &  \\ 
51 & 1.65      & M7.1 & M7 & 2700 & -2:   & 4.5 & Keck (low s/n) &  \\ 
\hline
\end{tabular}



%% file: 6308tab5.tex
\begin{tabular}{lllccc}
\hline
Assumed & magnitude & mass bin & number of & nb of & nb of \\
age & bin ($I$) & ($\msun$)& candidates &  contaminants & 
probable members \\
\hline
& 18.5-19.0 & 0.090-0.075 & 14 & 2$\pm$2 & 12$\pm$3 \\
100 Myr & 19.0-20.0 & 0.075-0.056 & 15 & 1$\pm$1 & 14$\pm$2 \\ 
& 20.0-21.0 & 0.056-0.041 & 12 & 1$\pm$1 & 11$\pm$2 \\ 
& 21.0-22.2 & 0.041-0.032 & 29 & 10$\pm$4 & 19$\pm$5 \\
\hline
& 18.5-19.0 & 0.104-0.086 & 15 & 3$\pm$2 & 12$\pm$3 \\
150 Myr & 19.0-20.0 & 0.086-0.065 & 19 & 3$\pm$2 & 16$\pm$3 \\ 
& 20.0-21.0 & 0.065-0.047 & 12 & 0$\pm$1 & 12$\pm$2 \\ 
& 21.0-22.2 & 0.047-0.037 & 30 & 10$\pm$5 & 20$\pm$6 \\
\hline
\end{tabular}

%% file: 6308.bbl
\begin{thebibliography}{}

\bibitem[Abraham de Epstein \& Epstein(1985)]{abra} Abraham de
  Epstein, A.E \& Epstein, I., 1985, AJ, 90, 1211
\bibitem[Adams et al.(2002)]{adams02} Adams, T., Davies, M.~B.,
  Jameson, R.~F., \& Scally, A. 2002, MNRAS, 333, 547
\bibitem[Allard et al.(2001)]{allard} Allard, F., Hauschildt, P.~H.,
  Alexander, D.~R., Tamanai, A., \& Schweitzer, A.\ 2001, ApJ, 556,
  357
\bibitem[Allen et al. 2006]{allen} Allen, L., Megeath, S.T.,
  Gutermuth, R., Myers, P.C., Wolk, S., Adams, F.C., Muzerolle, J.,
  Young, E., Pipher, J.L., 2006, to appear in "Protostars and Planets
  V", astro-ph/0603096
\bibitem[Ballesteros et al. 2006]{bal} Ballesteros-Paredes, J.,
  Klessen, R.S., Mac Low, M.-M., Vazquez-Semadeni, E., 2006, to appear
  in "Protostars and Planets V", astro-ph/0603357
\bibitem[Baraffe et al.(1998)]{bcah98} Baraffe, I., Chabrier, G.,
  Allard, F., Hauschildt, P.H. 1998, A\&A, 337, 403
\bibitem[Barrado et al.(2002)]{aper} Barrado y Navascu\'es, D.,
  Bouvier, J., Stauffer, J.R., Lodieu, N., McCaughrean, M.J. 2002,
  A\&A, 395, 813
\bibitem[Barrado y Navascu{\'e}s et al.(2004a)]{2004ApJ...614..386B} Barrado 
y Navascu{\'e}s, D., Stauffer, J.~R., \& Jayawardhana, R.\ 2004a, \apj, 614, 
386 
\bibitem[Barrado y Navascu{\'e}s et al.(2004b)]{2004ApJ...610.1064B} Barrado 
y Navascu{\'e}s, D., Stauffer, J.~R., Bouvier, J., Jayawardhana, R., \& 
Cuillandre, J.-C.\ 2004b, \apj, 610, 1064
\bibitem[Bate(2005)]{2005MNRAS.363..363B} Bate, M.~R.\ 2005, \mnras, 363, 
363 
\bibitem[Bate \& Bonnell(2005)]{2005MNRAS.356.1201B} Bate, M.~R., \& 
Bonnell, I.~A.\ 2005, \mnras, 356, 1201 
\bibitem[Bertin \& Arnouts(1996)]{ber} Bertin, E., \& Arnouts,
  S. 1996, A\&AS, 117, 393
\bibitem[B{\'e}jar et al.(2004)]{2004Ap&SS.292..339B} B{\'e}jar, V.~J.~S., 
Caballero, J.~A., Rebolo, R., Zapatero Osorio, M.~R., \& Y Navascu{\'e}s, 
D.~B.\ 2004, \apss, 292, 339 
\bibitem[Blanco(1949)]{blanco} Blanco, V.M., 1949, PASP, 61, 183
\bibitem[Bonnell et al. 2006a]{bon} Bonnell, I.A., Larson, R.B.,
  Zinnecker, H., 2006a, to appear in "Protostars and Planets V",
  astro-ph/0603447
\bibitem[Bonnell et al.(2006b)]{2006MNRAS.368.1296B} Bonnell, I.~A., Clarke, 
C.~J., \& Bate, M.~R.\ 2006b, \mnras, 368, 1296 
\bibitem[Brice{\~n}o et al.(2002)]{2002ApJ...580..317B} Brice{\~n}o, C., 
Luhman, K.~L., Hartmann, L., Stauffer, J.~R., \& Kirkpatrick, J.~D.\ 2002, 
\apj, 580, 317 
\bibitem[Chabrier(2003)]{2003PASP..115..763C} Chabrier, G.\ 2003, \pasp, 
115, 763 
\bibitem[Chabrier et al.(2000)]{cbah00} Chabrier, G., Baraffe, I.,
  Allard, F., Hauschildt, P.H. 2000, ApJ, 542, 464
\bibitem[Cuillandre et al.(2001)]{cui} Cuillandre, J.-C., Starr, B.,
  Isani, S., Lupino, G. 2001, Experimental Astronomy, v. 11, Issue 3,
  p. 223-235
\bibitem[de Wit et al.(2006)]{2006A&A...448..189D} de Wit, W.~J., et al.\ 
  2006, \aap, 448, 189
\bibitem[Dobbie et al.(2002)]{dobbie02} Dobbie, P.D., Pinfield,
  D.J., Jameson, R.F., \& Hodgkin, S.T.\ 2002, MNRAS, 335, L79
\bibitem[Eggen(1972)]{eggen} Eggen, O.J., 1972, ApJ, 173, 63
\bibitem[Epstein(1968)]{epst} Epstein, I., 1968, AJ, 73, 556
\bibitem[Ford et al.(2005)]{2005MNRAS.364..272F} Ford, A., Jeffries, R.~D., 
\& Smalley, B.\ 2005, \mnras, 364, 272 
\bibitem[Goodwin et al.(2006)]{2006A&A...452..487G} Goodwin, S.~P.,
Whitworth, A.~P., \& Ward-Thompson, D.\ 2006, \aap, 452, 487
\bibitem[Guieu et al.(2006)]{2006A&A...446..485G} Guieu, S., Dougados, C., 
Monin, J.-L., Magnier, E., \& Mart{\'{\i}}n, E.~L.\ 2006, \aap, 446,
485 
\bibitem[Hawkins \& Favata(1998)]{1998AAS...193.6707H} Hawkins, G.~W.,
  \& Favata, F.\ 1998, Bulletin of the American Astronomical Society,
  30, 1346
\bibitem[Jappsen et al.(2005)]{2005A&A...435..611J} Jappsen, A.-K., 
Klessen, R.~S., Larson, R.~B., Li, Y., \& Mac Low, M.-M.\ 2005, \aap, 435, 
611 
\bibitem[Jeffries \& James(1999)]{jj} Jeffries, R.~D.~\& James, D.~J.\
  1999, ApJ, 511, 218
\bibitem[Jeffries et al.(2004)]{2004MNRAS.351.1401J} Jeffries, R.~D., 
Naylor, T., Devey, C.~R., \& Totten, E.~J.\ 2004, \mnras, 351, 1401 
\bibitem[Kharchenko et al.(2005)]{2005A&A...438.1163K} Kharchenko,
N.~V., Piskunov, A.~E., R{\"o}ser, S., Schilbach, E., \& Scholz,
R.-D.\ 2005, \aap, 438, 1163
\bibitem[King(1962)]{king} King, I. 1962, AJ, 67, 471
\bibitem[King(1980)]{1980IAUS...85..139K} King, I.~R.\ 1980, IAU Symp.~ 85: 
Star Formation, 85, 139
\bibitem[Klessen et al.(2005)]{2005ApJ...620..786K} Klessen, R.~S., 
Ballesteros-Paredes, J., V{\'a}zquez-Semadeni, E., \& Dur{\'a}n-Rojas, C.\ 
2005, \apj, 620, 786 
\bibitem[Larson(2005)]{2005MNRAS.359..211L} Larson, R.~B.\ 2005, \mnras, 
359, 211 
\bibitem[Li et al.(2007)]{2007ApJ...655..351L} Li, D., Velusamy, T.,
  Goldsmith, P.~F., \& Langer, W.~D.\ 2007, \apj, 655, 351
\bibitem[Luhman et al.(2003a)]{2003ApJ...593.1093L} Luhman, K.~L., Stauffer, 
J.~R., Muench, A.~A., Rieke, G.~H., Lada, E.~A., Bouvier, J., \& Lada, 
C.~J.\ 2003a, \apj, 593, 1093 
\bibitem[Luhman et al.(2003b)]{2003ApJ...590..348L} Luhman, K.~L., 
Brice{\~n}o, C., Stauffer, J.~R., Hartmann, L., Barrado y Navascu{\'e}s, 
D., \& Caldwell, N.\ 2003b, \apj, 590, 348
\bibitem[Lyng{\aa} \& Wramdemark(1984)]{lynga} Lyng\aa, G. \&
  Wramdemark, S., 1984, A\&A, 132, 54
\bibitem[Lyng{\aa} (1987)]{lynga87} Lyng\aa, G. 1987, Catalogue of
  open cluster data. Computer based catalogue available through CDS,
  Strasbourg, France and through NASA Data Center, Greenbell,
  Maryland, USA.
\bibitem[Magnier et al.(2002)]{gene} Magnier, E., \& Cuillandre,
  J.-C. 2002, SPIE, 4844, 343
\bibitem[Magnier \& Cuillandre(2004)]{2004PASP..116..449M} Magnier, E.~A., 
\& Cuillandre, J.-C.\ 2004, \pasp, 116, 449 
\bibitem[Mart{\'{\i}}n et al.(1999)]{martin99} Mart{\'{\i}}n, E.~L.,
  Delfosse, X., Basri, G., Goldman, B., Forveille, T., \& Zapatero
  Osorio, M.~R.\ 1999, AJ, 118, 2466
\bibitem[Micela et al.(1999)]{micela} Micela, G., Sciortino, S.,
  Favata, F., Pallavicini, R., \& Pye, J., 1999, A\&A, 344, 83
\bibitem[Moraux et al.(2001)]{mor01} Moraux, E., Bouvier, J., Stauffer,
  J.R. 2001, A\&A, 367, 211
\bibitem[Moraux et al.(2003)]{mor03} Moraux, E., Bouvier, J., Stauffer,
  J.R., Cuillandre, J.-C.  2003, A\&A, 400, 891
\bibitem[Moraux et al.(2004)]{2004A&A...426...75M} Moraux, E., Kroupa, P., 
\& Bouvier, J.\ 2004, \aap, 426, 75 
\bibitem[Motte \& Andr{\'e}(2001)]{2001A&A...365..440M} Motte, F., \& 
Andr{\'e}, P.\ 2001, \aap, 365, 440 
\bibitem[Muench et al.(2002)]{2002ApJ...573..366M} Muench, A.~A., Lada, 
E.~A., Lada, C.~J., \& Alves, J.\ 2002, \apj, 573, 366 
\bibitem{nak} Nakajima, T., Oppenheimer, B.~R., Kulkarni, S.~R.,
  Golimowski, D.~A., Matthews, K., \& Durrance, S.~T.\ 1995, Nature,
  378, 463
\bibitem[Nutter \& Ward-Thompson(2007)]{2007MNRAS.374.1413N} Nutter,
  D., \& Ward-Thompson, D.\ 2007, \mnras, 374, 1413
\bibitem[Padoan \& Nordlund(2002)]{pn} Padoan, P., \& Nordlund, \AA.,
  2002, ApJ, 576, 870
\bibitem[Panagi \& O'dell(1997)]{1997A&AS..121..213P} Panagi, P.~M.,
  \& O'dell, M.~A.\ 1997, \aaps, 121, 213
\bibitem[Perry et al.(1978)]{perry} Perry, C.L., Walter, D.K.,
  Crawford, D.L., 1978, PASP, 90, 81
\bibitem[Persson et al.(1998)]{1998AJ....116.2475P} Persson, S.~E., Murphy, 
D.~C., Krzeminski, W., Roth, M., \& Rieke, M.~J.\ 1998, \aj, 116, 2475 
\bibitem[Pillitteri et al.(2003)]{pilli} Pillitteri, I., Micela, G.,
  Sciortino, S., \& Favata, F.\ 2003, A\&A, 399, 919
\bibitem[Pinfield et al.(1998)]{pin98} Pinfield, D.J., Jameson, R.F.,
  \& Hodgkin, S.T. 1998, MNRAS, 299, 955
\bibitem[Rebolo et al.(1995)]{1995Natur.377..129R} Rebolo, R.,
  Zapatero-Osorio, M.~R., \& Martin, E.~L.\ 1995, \nat, 377, 129
\bibitem[Robichon et al.(1999)]{1999A&A...345..471R} Robichon, N.,
  Arenou, F., Mermilliod, J.-C., \& Turon, C.\ 1999, \aap, 345, 471
\bibitem[Silverman(1986)]{1986desd.book.....S} Silverman, B.~W.\ 1986, 
Monographs on Statistics and Applied Probability, London: Chapman and Hall, 
1986
\bibitem[Spitzer(1940)]{1940MNRAS.100..396S} Spitzer, L.~J.\ 1940, \mnras, 
100, 396 
\bibitem[Westerlund(1963)]{west63} Westerlund, B.E., 1963, MNRAS, 127,
  183
\bibitem[Westerlund et al.(1988)]{west88} Westerlund, B.E., Garnier,
  R., Lundgren, K., Petterson, B., Breysacher, J., 1988, A\&AS, 76,
  101
\bibitem[Zheng et al.(2004)]{2004ApJ...601..500Z} Zheng, Z., Flynn, C., 
Gould, A., Bahcall, J.~N., \& Salim, S.\ 2004, \apj, 601, 500 


\end{thebibliography}
